\documentclass[a4paper,fleqn, review]{cas-sc}

\usepackage[authoryear]{natbib}

\usepackage{makecell}
\usepackage{multirow}
\usepackage{rotating}
\usepackage{array}
\usepackage{longtable}

\newcommand{\moving}{M}
\newcommand{\fixed}{F}
\newcommand{\jacdet}{det\mathbf{J} }
\newcommand{\jac}{\mathbf{J} }

\newcommand{\disp}{\mathbf{u} }
\newcommand{\velocity}{\mathbf{v} }

\usepackage{enumitem}
\usepackage{xcolor}
\definecolor{implcolor}{rgb}{0.64, 0.87, 0.93}
\definecolor{explcolor}{rgb}{0.27, 0.42, 0.81}
\definecolor{guidcolor}{rgb}{0.52, 0.73, 0.4}
\definecolor{dlcolor}{rgb}{0.91, 0.45, 0.32}
\definecolor{convcolor}{rgb}{0.88, 0.66, 0.37}
\definecolor{convcolor}{rgb}{1.0, 0.8, 0.2}
\newcommand{\explbox}{\fcolorbox{white}{explcolor}{\rule{0pt}{0.1em}\rule{0.1em}{0pt}}}
\newcommand{\implbox}{\fcolorbox{white}{implcolor}{\rule{0pt}{0.1em}\rule{0.1em}{0pt}}}
\newcommand{\guidbox}{\fcolorbox{white}{guidcolor}{\rule{0pt}{0.1em}\rule{0.1em}{0pt}}}
\newcommand{\dlbox}{\fcolorbox{white}{dlcolor}{\rule{0pt}{0.1em}\rule{0.1em}{0pt}}}
\newcommand{\convbox}{\fcolorbox{white}{convcolor}{\rule{0pt}{0.1em}\rule{0.1em}{0pt}}}

\def\tsc#1{\csdef{#1}{\textsc{\lowercase{#1}}\xspace}}
\tsc{WGM}
\tsc{QE}

\begin{document}
\let\WriteBookmarks\relax
\def\floatpagepagefraction{1}
\def\textpagefraction{.001}

\shorttitle{Regularization in
Medical Image Registration: A Comprehensive
Review}    

\shortauthors{Reithmeir, Spieker, Sideri-Lampretsa, Rueckert, Schnabel, Zimmer}  

\title [mode = title]{From Model Based to Learned Regularization in Medical Image Registration: A Comprehensive Review}  



%

\author[1,2,3]{Anna Reithmeir}[orcid=0009-0007-4449-3627, auid=1]
\cormark[1]
\cortext[1]{Corresponding author. E-mail address: anna.reithmeir@tum.de}

\author[1,3]{Veronika Spieker}[orcid=0000-0001-7720-7569]
\author[1]{Vasiliki Sideri-Lampretsa}[orcid=0000-0003-0135-7442]
\author[1,2,4,6]{Daniel Rueckert}[orcid=0000-0002-5683-5889]
\author[1,2,3,5]{Julia A. Schnabel}[orcid=0000-0001-6107-3009]
\author[1,6]{Veronika A. Zimmer}[orcid=0000-0002-5093-5854]





\affiliation[1]{organization={School of Computation, Information and Technology, Technical University of Munich (TUM)},
city={Munich},
country={Germany}}
\affiliation[2]{organization={Munich Center for Machine Learning (MCML)},
city={Munich},
country={Germany}}
\affiliation[3]{organization={Institute of Machine Learning in Biomedical Imaging, Helmholtz Munich},
city={Munich},
country={Germany}}
\affiliation[4]{organization={Department of Computing, Imperial College London},
city={London},
country={United Kingdom}}
\affiliation[5]{organization={School of Biomedical Engineering and Imaging Sciences, King’s College London},
city={London},
country={United Kingdom}}
\affiliation[6]{organization={School of Medicine, Klinikum rechts der Isar,
Technical University of Munich},
city={Munich},
country={Germany}}






\nonumnote{Funding: A.R. is partially supported by the EVUK programme "Next-generation Al for Integrated Diagnostics” of the Free State of Bavaria; V. S.-L. is partially supported by ERC Grant Deep4MI (Grant No. 884622); V.S. is partially supported by the Helmholtz Association under the joint research school “Munich School for Data Science - MUDS”.} 


\begin{abstract}
Image registration is fundamental in medical imaging applications, such as disease progression analysis or radiation therapy planning. 
The primary objective of image registration is to precisely capture the deformation between two or more images, typically achieved by minimizing an optimization problem.
Due to its inherent ill-posedness, regularization is a key component in driving the solution toward anatomically meaningful deformations. A wide range of regularization methods has been proposed for both conventional and deep learning-based registration. However, the appropriate application of regularization techniques often depends on the specific registration problem, and no ’one-fits-all’ method exists.
Despite its importance, regularization is often overlooked or addressed with default approaches, assuming existing methods are sufficient. A comprehensive and structured review remains missing.
This review addresses this gap by introducing a novel taxonomy that systematically categorizes the diverse range of proposed regularization methods. It highlights the emerging field of \textit{learned regularization}, which leverages data-driven techniques to automatically derive deformation properties from the data. 
Moreover, this review examines the transfer of regularization methods from conventional to learning-based registration, identifies open challenges, and outlines future research directions. By emphasizing the critical role of regularization in image registration, we hope to inspire the research community to reconsider regularization strategies in modern registration algorithms and to explore this rapidly evolving field further. 
\end{abstract}


\maketitle

\section{Introduction}

Image registration is crucial in many clinical applications, e.g., in radiation therapy or disease monitoring \citep{rueckert2011, sotiras2013}. 
Its primary objective is to accurately identify the deformation between two or more images, typically by minimizing an optimization problem. 
In recent years, learning-based registration methods have become state-of-the-art. Unlike conventional methods that iteratively solve an optimization problem for each image pair, learning-based approaches use neural networks to parameterize the transformation model and to learn the registration with training data. Once trained, these models can register unseen image pairs in real time. 

A key challenge in image registration lies in its ill-posed nature: Mathematically, multiple solutions exist, but only few are anatomically or physiologically feasible.
Achieving such plausible solutions is crucial for accurate intra-patient registration in clinical settings, where the deformation should reflect the true motion of the physical structure.
For instance, when deformation causes image regions to overlap, this is called folding (see Fig. \ref{fig:smo}), which is physically impossible for soft tissue and thus should not be present in its deformation.

The key to achieving a unique and realistic solution to a registration problem is the regularization of the optimization problem. 
Regularization constrains the solution space and incorporates deformation characteristics into the process. It is an essential element in almost every registration algorithm, forming a \textit{fundamental building block} of a successful registration algorithm -- alongside the (dis-)similarity measure, transformation model, optimization procedure, and validation protocol as outlined in \citep{rueckert2011}.\\

A wide variety of regularization methods has been proposed with different aims. These range from enforcing general properties such as spatial smoothness and invertibility to addressing more complex ones, including modeling organ specific motion or handling missing regions. 
While most of these methods originate in conventional registration, several have already been successfully adapted to learning-based registration \citep{learn2reg2022}.
However, despite the widely acknowledged importance of regularization, its role is often neglected in practice, and many state-of-the-art registration frameworks rely on standard techniques, such as $\mathcal{L}_2$-norm smoothing, which may not fully address real-world deformations.
Furthermore, regularization approaches often seem handcrafted, and a 'one-fits-all' method does not exist.\\

\begin{figure}
    \centering
    \includegraphics[width=.9\textwidth]{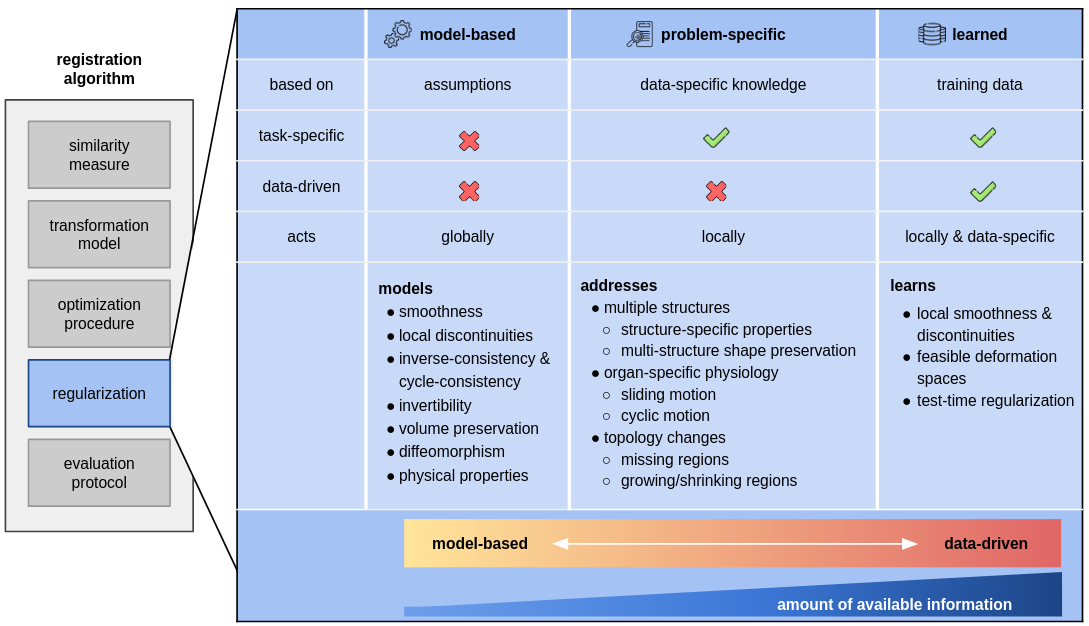}
    \caption{Regularization is an essential building block of successful registration algorithms. We identify three main categories of regularization methods: (I) Model based regularization that imposes prior assumptions on the deformation; (II) problem specific regularization that takes into account additional knowledge about the data, such as spatial information in the form of segmentation maps or physiological information; and (III) Learned regularization, which derives deformation properties from training data with a machine or deep learning model. The amount of prior information included
in the regularization increases from category I to III. Most problem specific and learned regularization methods have their origin in model based regularization.
    }
    \label{fig:overview}
\end{figure}

The lack of a comprehensive taxonomy that organizes the diverse landscape of regularization techniques hinders the potential method transfer across different image registration applications and limits progress in the field.
Several general review papers exist on conventional and deep learning-based medical image registration methods \citep{viergever2016, haskins2020review, fu2020review, chen2023survey} as well as targeted reviews on specific registration building blocks like deformation models \citep{wang2019review} or registration error estimation \citep{bierbrier2022}, and application-specific areas \citep{ferrante2017, matl2017}. While some of these reviews briefly cover prominent regularization methods, no structured review exists that fully explores the many proposed regularization techniques in both conventional and learning-based medical image registration algorithms. 

In this review, we address this gap by providing the first comprehensive overview of regularization techniques in conventional and deep learning-based image registration. 
We identify three main categories (see also Fig. \ref{fig:overview}):
Regularization, which is (I) model based and takes prior assumptions into account
(Sec. \ref{sec:prior-assumptions}), (II) problem specific and incorporates additional knowledge 
(Sec. \ref{sec:prior-knowledge}), and (III) learned from a training dataset using machine or deep learning models (Sec. \ref{sec:learned-knowledge}). 

The goals of this review are three-fold. First, we intend to give researchers a structured overview of existing regularization techniques. 
Second, we aim to facilitate the transfer of regularization methods across different registration techniques and applications. 
To this end, we examine to what extent regularization techniques have been transferred to learning-based registration and highlight the emerging field of learned regularization. Moreover, we discuss open challenges and promising future directions.
Finally, we hope to inspire the research community to further explore regularization in deep learning-based image registration, encouraging innovation in this rapidly evolving field. 

The review includes articles published up to October 2024 that contribute novel methodologies for regularization in pairwise medical image registration.
We searched for papers in Scopus, PubMed and GoogleScholar using combinations of the keywords "medical image registration", "regulariz(s)ation", "regulariz(s)er", and the terms presented in Fig. \ref{fig:overview}.
As our focus lies on the regularization aspect, the proposed taxonomy targets the \textit{regularization} only, independent from the \textit{registration} (where possible). 
We do not intend to give an overview of deep learning-based registration methods (refer to the reviews mentioned above). 
However, for a better overview, we mention whether the regularization techniques were initially proposed for conventional or learning-based registration.

The remainder of this review is structured as follows:
\begin{itemize}
\item \textbf{Section 2}: Brief background on medical image registration and regularization. The mathematical notation used in this review is presented here;
\item \textbf{Section 3}: Presentation and discussion of the three main categories of regularization methods found in the literature;
The method transfer from conventional to learning-based registration is also examined;
\item \textbf{Section 4}: Discussion of open challenges and future directions in the field;
\item \textbf{Section 5}: Concluding remarks.
\end{itemize}

\section{Background}
\label{sec:bg}
In this section, we cover the relevant mathematical background, notation, and definitions for image registration, as well as its regularization. \\

\textbf{Image registration:} Given two images $\moving{}, \fixed{}: \Omega \subset \mathbb{R}^{N}\rightarrow\mathbb{R}$ with $N\in\{2,3\}$, deformable image registration aims to find an optimal spatial deformation $\phi: \mathbb{R}^N \rightarrow \mathbb{R}^N $, so that $ \fixed{}\approx \moving{} \circ \hspace{1pt} \phi$. Generally, this deformation is obtained by minimizing the following optimization problem:
\begin{equation}
    \label{eq:opt}
    \mathcal{L}(\fixed{}, \moving{}, \phi) = \mathcal{L}_{Sim}(\fixed{}, \moving{} \circ \hspace{1pt} \phi) +\alpha \mathcal{L}_{Reg}(\phi).
\end{equation}
Here, $\mathcal{L}_{Sim}$ is a dissimilarity measure that quantifies how well the deformed moving image \( \moving{} \circ  \phi \) matches the fixed image \( \fixed{}\). 
Common choices for \( \mathcal{L}_{Sim} \) include the sum of squared distances (SSD) and the negative normalized cross-correlation (NCC) \citep{haskins2020review}. 
The regularization term $\mathcal{L}_{Reg}$ constrains the solution space. This is necessary since the optimization problem is inherently ill-posed, i.e., multiple solutions can exist. 
With this term, the registration can be driven towards a solution with desired properties. 

\begin{figure}
    \centering
    \includegraphics[width=0.6\textwidth]{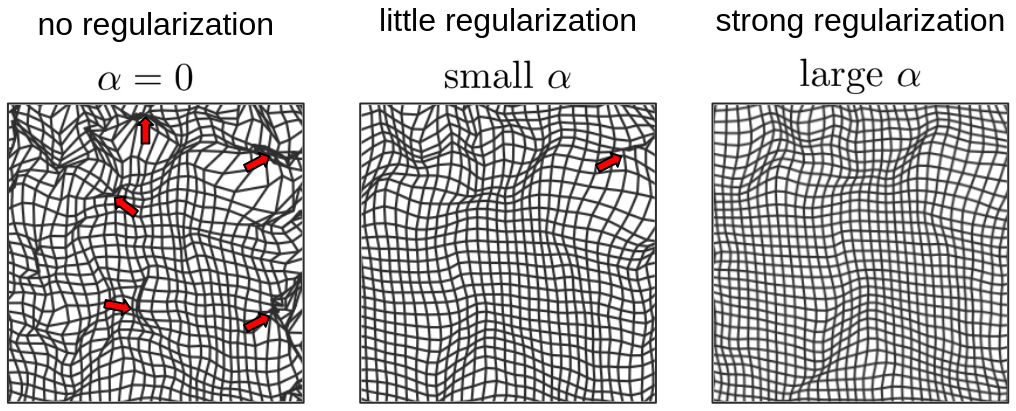}
    \caption{Model based regularization: Smoothness and folding: Different levels of smoothness regularization, controlled with the regularization parameter $\alpha$ (see Eq. \ref{eq:opt}). The pink arrows indicate regions of folding. With increasing $\alpha$, more smoothing is applied and less folding is observed. }
    \label{fig:smo}
\end{figure}

Both terms are balanced by the regularization weight $\alpha\in\mathbb{R}^+$, which is a hyperparameter that is typically tuned manually. 
For high values of $\alpha$, the deformation properties are given more importance than the image alignment. In contrast, for low values, the intensity alignment is given more weight, and the deformation is more flexible (see Fig. \ref{fig:smo} for an example).
For more details on medical image registration in general, see, e.g., \citep{rueckert2011}.\\

\textbf{The displacement field Jacobian and its determinant:}
In deformable image registration, the spatial deformation $\phi$ is typically parameterized as a discretized displacement field $\disp\in\mathbb{R}^{H\times W(\times D) \times N}$ where a displacement vector is assigned to each voxel position $x$. 
The Jacobian matrix $\jac$ of $\disp$ is defined as 
\begin{equation}
    \jac{}= \begin{bmatrix}
\frac{\partial u_x}{\partial x} & \frac{\partial u_x}{\partial y} & \frac{\partial u_x}{\partial z} \\
\frac{\partial u_y}{\partial x} & \frac{\partial u_y}{\partial y} & \frac{\partial u_y}{\partial z} \\
\frac{\partial u_z}{\partial x} & \frac{\partial u_z}{\partial y} & \frac{\partial u_z}{\partial z}
\end{bmatrix} \in \mathbb{R}^{H\times W (\times D)\times N\times N}
\end{equation}
and contains the first-order partial derivatives of $\disp$.
An important concept is the Jacobian determinant $\jacdet{}\in\mathbb{R}^{H\times W(\times D)}$. It indicates local volume and orientation change within a small neighborhood of $x$:
\begin{equation}
\label{eq:jacdet}
\text{local}
\begin{cases}
\text{tearing } & \text{if } \jacdet(x) \to\infty, \\
\text{volume increase } & \text{if } \jacdet(x) > 1, \\
\text{volume preservation } & \text{if } \jacdet(x) = 1, \\
\text{volume decrease } & \text{if } \jacdet(x) < 1, \\
\text{singularity/folding } & \text{if } \jacdet(x) < 0.
\end{cases}
\end{equation}
Folding means that after the deformation is applied, multiple image regions overlap, as visualized in Fig. \ref{fig:smo}. In this case, the deformation locally lacks invertibility and orientation preservation. 
The Jacobian determinant plays an important role not only in the analysis of deformation fields, e.g., to assess their smoothness, but also for the regularization of registration techniques.\\

\begin{figure}
    \centering
    \includegraphics[width=\textwidth]{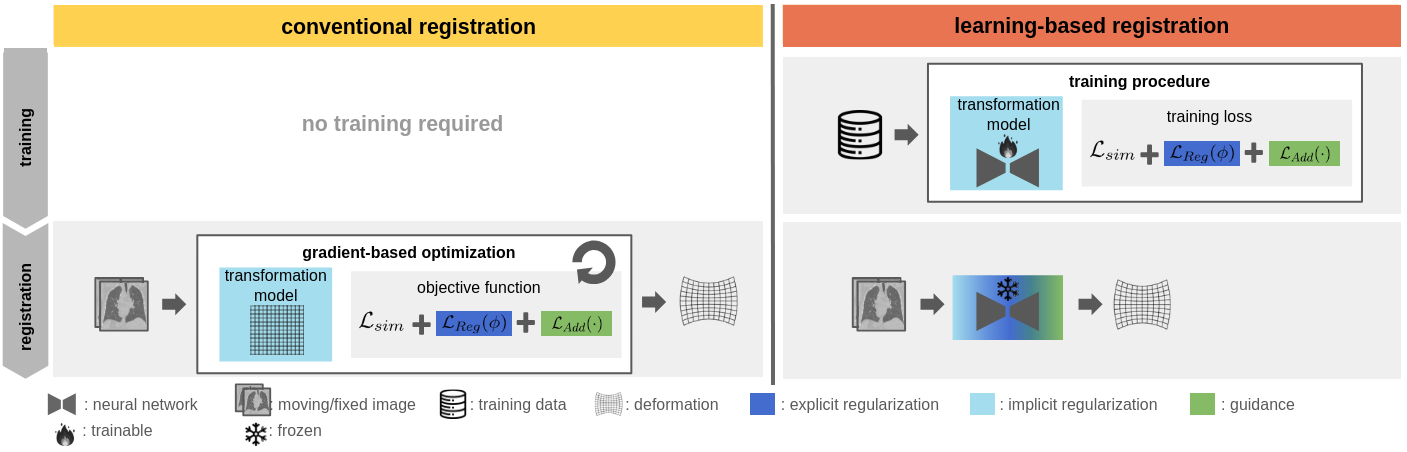}
    \caption{
    Explicit vs. implicit regularization: Overview of approaches to integrating regularization in conventional (left) and learning-based (right) medical image registration.
    In both, regularization can be achieved explicitly with a regularizing loss term (dark blue) or implicitly with the parameterization of the transformation model (light blue). Additionally, guiding loss terms (green) can drive the registration toward a desired solution.
    For learning-based registration, the regularization applied during training is inherently captured in the network parameters at inference time.
    }
    \label{fig:expl-impl}
\end{figure}

\textbf{Explicit vs. implicit regularization and guidance:}
We differentiate two kinds of regularization:
\begin{itemize}[noitemsep,nolistsep]
    \item \textit{Explicit regularization}, which operates on the deformation and comes in the form of a loss term, and
    \item \textit{Implicit regularization}, which arises from the parameterization of the deformation model.
\end{itemize}
In conventional registration, explicit regularization is achieved by a regularizing loss term $\mathcal{L}_{Reg}(\phi)$ in the objective function Eq. \ref{eq:opt} (Fig. \ref{fig:expl-impl}, left). 
In contrast, implicit regularization is achieved, for instance, with the B-Spline-based transformation parameterization in free-form deformations (FFD) \citep{rueckert1999} and coarse-to-fine multiresolution approaches, as in \citep{mok2020lapirn}. They naturally enforce smoothness without requiring explicit constraints. 
In learning-based registration, explicit and implicit  regularization is applied during network training (Fig. \ref{fig:expl-impl}, top right). 
At inference time, the regularization is inherently incorporated in the trained network weights. (Fig. \ref{fig:expl-impl}, bottom right).
In the context of learning-based registration, this review focuses on regularization strategies that constrain the deformation space during training.

Beyond explicit regularization loss terms, additional \textit{guiding loss terms} $\mathcal{L}_{Add}(\cdot)$, such as segmentation overlap measures \citep{balakrishnan2019voxelmorph}, can indirectly encourage plausible solutions and drive the registration toward a desired result. 
Since such guiding terms do not operate on the deformation field, they are, strictly speaking, not a type of regularization. However, we believe that guidance approaches can be equally important in obtaining plausible registration solutions as "real" regularization, and thus include some widely adopted methods in this review.

\section{Regularization Techniques in Medical Image Registration}
Within the medical image registration literature, three main categories regarding regularization methods can be identified:
\begin{itemize}
\item Category I: \textit{Model based regularization with prior assumptions} (Sec. \ref{sec:prior-assumptions}).
This type of regularization models user-defined assumptions about the deformation properties, such as smoothness, inverse-consistency, invertibility, diffeomorphisms, volume preservation, or based on physical principles. It is applied globally, i.e., the same regularization is applied similarly at every location of the image. 
\item Category II: \textit{problem specific regularization with prior data knowledge} (Sec. \ref{sec:prior-knowledge}).
This type of regularization enriches the registration process with a priori available knowledge about the data. 
On the one hand, this can be spatial information about the image content, which allows for modeling structure specific deformation properties and preserving intra-organ topology. On the other hand, this can be information about the clinical context of the data or the physiology of the organ in focus. 
For example, the knowledge that the data consists of pre- and post-operative images indicates the presence of missing regions, and the knowledge that the images are inhale-exhale lung image pairs suggests to model sliding motion of the lung.
Problem specific regularization is spatially adaptive, meaning that it depends on the image location and can adapt to local properties.
\item Category III: \textit{Learned regularization} (Sec. \ref{sec:learned-knowledge}). This type of regularization is data-driven and learns the deformation properties directly from a training dataset. It is often spatially adaptive and is learned either independently before being applied in the registration or jointly alongside the registration. Learned regularization is typically parameterized as a deep learning model. 
\end{itemize}
As becomes clear, the amount of prior information included in the regularization increases from category I to III. 
An overview of the proposed taxonomy is shown in Fig. \ref{fig:overview}, and the three categories are presented in the following.

\subsection{Model Based Regularization with Prior Assumptions}
\label{sec:prior-assumptions}
In this section, we present and discuss model based regularization techniques, organized by the deformation properties they enforce. Model based regularization imposes a user-defined model on the deformation properties and is governed by prior assumptions, such as deformation invertibility or physical principles.
These regularization methods are uniformly applied across the image domain.
Tab. \ref{tab:modelbased} summarizes the most prominent model based techniques.

\subsubsection{Smoothness}
\label{sec:smo}
A spatially smooth deformation changes gradually across the image, where neighboring image locations deform similarly and no abrupt changes or folding are present (see Fig. \ref{fig:smo}). This property is particularly desirable, as most biological tissues deform smoothly.
Among the various methods for enforcing smoothness, explicit diffusion regularization is perhaps the most widely used method in image registration. It promotes smoothness by penalizing the $L_2$-norm of the spatial deformation gradient. 
Originating from the optical flow estimation of \cite{horn1981}, it is found in many state-of-the-art registration frameworks, including, e.g., VoxelMorph \citep{balakrishnan2019voxelmorph}.
Alternatively, smoothness regularization can be achieved using a Gaussian kernel, as demonstrated in the well-known Demons algorithm \citep{thirion}. 
While diffusion regularization considers first-order derivatives, curvature regularization \citep{fischer2004} leverages second-order derivatives to approximate the curvature of the displacement components. 
Recently, \cite{song2023} proposed using average pooling layers within registration networks to enforce smoothness, encouraging voxel displacements to align with the mean displacement in their neighborhood.

Smoothness can also be imposed implicitly through the transformation model.
For instance, free-form deformation (FFD) registration \citep{rueckert1999} parameterizes the deformation with cubic B-splines, which are inherently smooth. Extensively applied in conventional registration, FFD registration has also been successfully adapted to learning-based frameworks, such as in \cite{qiu2021bsplines}.
Moreover, multiresolution registration strategies, which derive high resolution deformations from coarser levels, impose smoothness implicitly and are widely regarded as state-of-the-art, particularly in learning-based registration. Training strategies range from separately training each resolution level \citep{devos2019, hering2019, mok2020lapirn,gu2021} to end-to-end training across all levels \citep{fu2018, wodzinski2021, yang2022, ma2023}. Beyond image resolution, multiresolution schemes have been extended to feature resolution \cite{wang2023modet}.

For spatio-temporal 4D (3D+t) data, smoothness can be extended to the temporal dimension. Examples include spatio-temporal Gaussian filters \citep{shen2005}, B-spline transformation models \citep{perperidis2005}, and multiresolution schemes \citep{aggrawal2020}.

\subsubsection{Invertibility}
\label{sec:inv}
Most biological tissue deformations observed in the human body are inherently reversible. 
According to the inverse function theorem, a deformation is locally invertible if the Jacobian determinant $\jacdet$ is positive, indicating the absence of folding. 
One of the first methods to ensure local invertibility involves monitoring $\jacdet$ throughout the registration process and re-gridding the image whenever it approaches zero, followed by restarting the optimization \citep{christensen1996}.
Explicit invertibility regularization includes applying penalties to areas that locally collapse to zero volume \citep{edwards1998}, to very small values of $\jacdet$ \citep{Christensen2001}, or constraining $\jacdet$ to remain within a pre-defined positive range \citep{haber2007}.

Invertibility of FFD B-spline deformations can be achieved by constraining $\jacdet$ at the control points to be above a positive threshold while penalizing large displacement derivatives between control point locations \citep{sdika2008}. Alternatively, the difference between neighboring spline coefficients can be constrained, as proposed in \cite{chun2008}.
Recently, \cite{huang2024} introduced a topology-preserving regularization using Beltrami coefficients, which quantify local distortions.

Explicit invertibility regularization has also been effectively incorporated into learning-based registration. For instance, folding can be mitigated by applying a rectified linear unit (ReLU) activation function to $-\jacdet$ \citep{mok2021, zhu2022} or by penalizing the sum of negative $-\jacdet$ values within the training loss function \citep{estienne2021}. 
\cite{pal2022} eliminate folding by reconstructing the predicted deformation from the matrix exponential of its Jacobian. This reconstruction inherently ensures positive $\jacdet$, and the reconstruction loss is seamlessly integrated as an explicit regularization term in the training loss function.

\subsubsection{Inverse-Consistency and Cycle-Consistency}
\label{sec:ic}
\begin{figure}[t]
    \centering
    \includegraphics[width=0.35\textwidth]{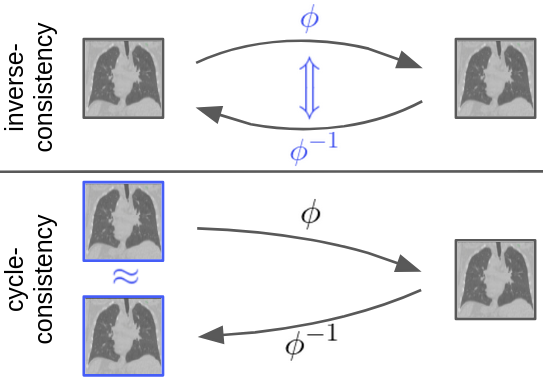}
    \caption{Model based regularization -- Inverse- vs. cycle-consistency: Inverse-consistency ensures that the forward and backward deformations are inverses of each other. Cycle-consistency ensures that a forward-backward deformed image resembles the original image. }
    \label{fig:ic-cc}
\end{figure}

The existence of an inverse deformation does not necessarily guarantee its plausibility. 
To ensure meaningful results, the forward deformation $\phi_{M\rightarrow F}$ (from $\moving{}$ to $\fixed{}$) should be the exact inverse of the backward deformation $\phi_{F\rightarrow M}$ (from $\fixed{}$ to $\moving{}$), making the registration independent of the order in which the two images are processed (see Fig. \ref{fig:ic-cc}). 
This property, called inverse-consistency (IC), can be explicitly enforced by penalizing deviations between the identity and the composed forward-backward deformation. Originally proposed for conventional registration in \cite{Christensen2001, leow2005}, it has since been adapted to learning-based registration \citep{zhang2018inverseconsistent, shen2019, estienne2021, greer2021}. 
Moreover, IC regularization can be found in symmetric registration, which simultaneously optimizes both registration directions, e.g., in the registration network of \cite{sha2024}.
While some symmetric methods calculate the IC loss with the inverses of the estimated deformations \citep{zhang2018inverseconsistent, shen2019}, \cite{leow2005} model the backward deformation by inverting the forward deformation to reduce computational effort.

Recent advancements extend inverse-consistency regularization to learning-based registration. \cite{zhang2023} introduce a robust IC loss that averages the predicted forward and inverted backward deformations, while \cite{duan2023saner} relax the IC constraint proportionally to the deformation magnitude.
The GradICON regularization \citep{tian2022} departs from previous methods by explicitly penalizing deviations of the \textit{gradient} of the forward-backward composition from the identity deformation. This maintains good IC properties while showing much better training convergence.
Multiresolution registration networks with built-in IC and symmetry properties have been recently proposed \citep{greer2023, honkamaa2024}.

While IC considers the deformation space, the same effect can be achieved by posing constraints in image space instead, a concept called cycle-consistency (CC). Here, an image should remain unchanged after the combined forward-backward deformation is applied (Fig. \ref{fig:ic-cc}). 
This is typically achieved with a guidance loss that penalizes deviations between the original and the forward-backward deformed image.
CC guidance is commonly incorporated in the training loss of a registration network, e.g., in \cite{lu2019, kim2021, nazib2021, huang2021, zhou2023}, and incorporated in advanced network architectures such as 
cycle-GANs \citep{mahapatra2018, zheng2022}, symmetric attention layers \citep{gao2022}, and cycle-consistent implicit neural representations \citep{vanharten2024}.

\subsubsection{Diffeomorphisms}
\label{sec:diff}
A diffeomorphic deformation is not only smooth but also constitutes a one-to-one mapping with a smooth inverse. 
As a homeomorphic transformation, it preserves the underlying topology, which is essential given that anatomical structures generally maintain their topology during deformation. As it equally ensures smoothness, invertibility, and topology preservation, diffeomorphism regularization can be useful.

A common approach for achieving diffeomorphisms is implicit regularization via the transformation model. 
The space of diffeomorphisms forms a Riemannian manifold, and the registration problem can be viewed as the search for the shortest geodesic (which can be considered as identifying the least distorted diffeomorphic deformation). This geometric optimization is typically formulated as ordinary differential equations (ODEs) based on smooth stationary velocity fields (SVF) \citep{Arsigny2006} or time-dependent velocity fields, as in the large deformation diffeomorphic metric mapping (LDDMM) \citep{Beg2005, Ashburner2011}.
To ensure the necessary SVF smoothness, the diffusion regularizer or constraints on $\jacdet$ can be additionally applied \citep{Dupuis1998, Beg2005}. 
The optimized SVF is then integrated over time to obtain the displacement field, e.g., with the scaling-and-squaring method \citep{Arsigny2006}. 

The optimization on the manifold of diffeomorphisms is computationally demanding. Thus, the log-Euclidean framework \citep{Arsigny2006} performs the optimization in the Euclidean tangent space of the manifold.
Several well-established conventional registration methods leverage diffeomorphisms, such as symmetric normalization (SyN) \citep{Avants2008}, which optimizes both images towards the mean image, and diffeomorphic Demons \citep{Vercauteren2009}. A geodesic shooting approach was proposed in \cite{Ashburner2011}, which optimizes only the initial momentum SVF and
an IC regularization within diffeomorphic registration is found in \cite{Beg2007ic}.

Diffeomorphic registration has been well transferred to learning-based registration. 
For instance, the registration networks in \cite{krebs2018, dalca2019, mok2020lapirn} predict SVFs and integrate them with a differentiable scaling-and-squaring integration layer. 
Learning-based adaptations have furthermore been developed for, e.g., the mean-image approach of SyN \citep{mok2021, ma2023} and geodesic shooting \citep{shen2019, ramon2024}.
Recent advancements include imposing inverse-consistency within diffeomorphic registration with a vision transformer architecture \citep{xu2024}, employing Lipschitz-continuous ResNet blocks to numerically approximate the flow field ODE \citep{joshi2022}, and leveraging time-embedded networks \citep{matinkia2024}.  

\subsubsection{Volume Preservation and Incompressibility} 
\label{sec:vp}
Many biological tissues are (nearly) incompressible, meaning they maintain their volume under pressure. 
This applies particularly to organs with high water or blood content, such as cardiac muscle \citep{bistoquet2008} and the liver \citep{hinkle2009}. 
Volume changes of deformations are quantified by the Jacobian determinant $\jacdet{}$ where perfect volume preservation is achieved for $\jacdet{}=1$. (see Eq. \ref{eq:jacdet}). A straightforward method to enforce volume preservation explicitly is to penalize deviations of $\jacdet{}$ from unity \citep{haber2004}. 
To equally penalize both expansion ($\jacdet>1$) and compression ($\jacdet<1$), \cite{Rohlfing2003} use the logarithm of $\jacdet$. 
Extensions to the above constraints include a relaxed volume preserving regularization that allows volume shrinking and growth within a certain range \citep{li2024} and spatiotemporal volume preservation constraints \citep{zhao2016}.

A different explicit approach to enforcing incompressibility is constraining deformations to be divergence-free. This implies volume preservation since the vector field divergence measures the extent to which each voxel behaves as a source. Divergence-free deformations can be achieved with the Helmholtz decomposition, which splits deformations into curl- and divergence-free components \citep{saddi2007, Mansi2011ilogd}. Alternatively, transformation parameterizations using divergence-free radial basis functions \citep{bistoquet2008} or divergence-conforming B-splines \citep{fidon2019} can be employed for implicit regularization. 

While incompressibility regularization is well studied in conventional registration, it is less frequently applied in learning-based registration. A notable exception is the physics-informed neural network (PINN, see Sec. \ref{sec:phy}) of \cite{ARRATIALOPEZ20235}, which penalizes deviations of $\jacdet$ from unity in the training loss function.

\begin{table}[t]
\centering
\caption{Model based regularization: Overview of model based regularization with prior assumptions about the deformation properties. The type of regularization (\explbox = explicit, \implbox = implicit, \guidbox = guidance) and registration framework in which the respective approach has been \textit{originally} proposed (\convbox = conventional, \dlbox = deep learning-based) are given. For simplicity, we omit $\int_{\Omega}\cdot\ dx$. Notation: exp=matrix exponential, 
$||\cdot||_F$ = Frobenius norm, 
$\partial_{ij}\mathbf{u}=\partial_{x_i}\mathbf{u}_j$, 
$\partial^2_{ij}\mathbf{u}=\partial^2_{x_ix_j}\mathbf{u}$, 
$\text{len}(\mathbf{u})$=$||\nabla \mathbf{u}-I||_F^2$,
cof=cofactor, 
$\text{surf}(\mathbf{u})$=$(||\text{cof} \nabla u||_F^2-3)^2$, $\text{vol}(\mathbf{u})=((\jacdet -1 )^2/\jacdet)^2$, $\text{div}(\mathbf{u})$= divergence component, $\text{curl}(\mathbf{u})$= curl component.}
\begin{scriptsize}
\begin{tabular}{llclcc}
\hline\hline
\textbf{\makecell[l]{deformation\\ property}}& \multicolumn{1}{l}{\textbf{\makecell[l]{reference}}}& \textbf{type}&\multicolumn{1}{l}{\textbf{\makecell[l]{approach}}} & \multicolumn{1}{l}{\textbf{\makecell[c]{regularization term}}} &\multicolumn{1}{c}{\textbf{\makecell[c]{registration\\framework}}} \\
 \hline
 \multirow{7}{*}{\makecell[c]{smoothness}}& \citep{horn1981}&\explbox& diffusion/$L_2$-norm &$\sum^N_i||\nabla\mathbf{u}_i||_2^2$&\convbox\\
& \citep{thirion}&\explbox& Gaussian kernel &&\convbox\\
& \citep{fischer2004}&\explbox& curvature &$\sum^N_i(\nabla\mathbf{u}^2_i)^2$&\convbox\\
& \citep{song2023}&\explbox& average pooling layer &&\dlbox\\
 &e.g. in \citep{mok2020lapirn}&\implbox& multiresolution registration& &\convbox\\
 & \citep{rueckert1999}& \implbox&B-spline FFD&&\convbox\\
 \hline
 \multirow{6}{*}{\makecell[c]{invertibility}}& \citep{Christensen2001}&\explbox&avoid negative $\jacdet$&$ (1/det\mathbf{J})^2$&\convbox\\
 & \citep{christensen1996}& \implbox&monitor $\jacdet$ \& re-grid&  &\convbox\\
 & \citep{mok2021}&\explbox&penalize negative $\jacdet$&$\text{ReLU}(-det\mathbf{J}$)&\dlbox\\
 & \citep{estienne2021}&\explbox&penalize negative $\jacdet$&$ \text{max}(0,-det\mathbf{J})$&\dlbox\\
 & \citep{pal2022}&\explbox&matrix exp. reconstruction&$\sum_i^N||\text{exp}(\jac)-\mathbf{J}_i||_2^2$&\dlbox\\
 \hline
 
 \multirow{4}{*}{\makecell[c]{inverse \\consistency}}&\citep{Christensen2001}&\explbox&forward-backward loss&$ ||\mathbf{u}_{M\rightarrow F} \circ \mathbf{u}_{F\rightarrow M}-\mathbf{I}||^2 $&\convbox\\
 & \citep{tian2022}&\explbox&GradICON&$ ||\nabla(\mathbf{u}_{M\rightarrow F} \circ \mathbf{u}_{F\rightarrow M})-\mathbf{I}||_F^2$&\dlbox\\
 & e.g. in \citep{lu2019}&\guidbox&cycle-consistency loss&$ || M - M\circ\mathbf{u}_{M\rightarrow F}\circ\mathbf{u}_{F\rightarrow M}||$ &\dlbox\\
 \hline
 
 \multirow{4}{*}{\makecell[c]{diffeomorphism}}&\citep{Arsigny2006}&\implbox&stationary velocity field (SVF)&&\convbox\\
 &\citep{Beg2005}&\implbox&time-dependent velocity field&&\convbox\\
  &\citep{Ashburner2011}&\implbox&momentum-based SVF&&\convbox\\
 \hline
 \multirow{6}{*}{\makecell[c]{volume \\preservation}}& \citep{Rohlfing2003}&\explbox&penalize $\jacdet \neq 1$& $||\text{log}(\jacdet)||$&\convbox\\
 &\citep{haber2004}&\explbox&penalize $\jacdet\neq 1$&$\jacdet-1$&\convbox\\
 &\citep{saddi2007}&\implbox&divergence-free deformations&&\convbox\\
 &\citep{bistoquet2008}&\implbox&divergence-free basis functions&&\convbox\\
 &\citep{fidon2019}&\implbox&divergence-conforming B-splines&&\convbox\\
 \hline
 
 \multirow{4}{*}{\makecell[c]{local \\discontinuities}}&\citep{chumchob2013}&\explbox& anisotropic TV/$L_1$-norm &$\sum^N_i||\nabla\mathbf{u}_i||_1$& \convbox\\
 &\citep{vishnevsky2017isotropic}&\explbox& isotropic TV &${\sqrt {\sum^N_{i,j}\nabla_i\mathbf{u}_j}}$& \convbox\\
 &\citep{nie2019}&\implbox&bounded deformations&&\convbox\\
 \hline
 
 & \citep{rueckert1999}&\explbox&bending energy&$ \sum^N_{i} (\partial^2_{ii}\mathbf{u})^2 +\sum^N_{i,j} 2({\partial^2_{ij}\mathbf{u}})^2
 $&\convbox\\
 \multirow{7}{*}{\makecell[c]{physics inspired}}& \citep{bookstein1989}&\implbox&bending energy&&\convbox\\
 &\citep{broit1981}&\explbox&linear elasticity &$ \frac{\mu}{4} \sum^N_{i,j}(\partial_{ij}\mathbf{u}+\partial_{ji}\mathbf{u})^2+\frac{\lambda}{2}(\text{div }\mathbf{u})^2$&\convbox\\
 &\citep{christensen1994}&\explbox&viscous fluid &$ \frac{\mu}{4} \sum^N_{i,j}(\partial_{ij}\mathbf{v}+\partial_{ji}\mathbf{v})^2+\frac{\lambda}{2}(\text{div }\mathbf{v})^2$&\convbox\\
  &\citep{burger2013}&\explbox&hyper-elasticity &$\text{len}(\mathbf{u}) + \text{surf}(\mathbf{u})+\text{vol}(\mathbf{u})$&\convbox\\
  &\citep{ARRATIALOPEZ20235}&\explbox&hyperelastic PINN &&\dlbox\\
  &\citep{yang2022}&\explbox&compressed regularization &&\dlbox\\
  &\citep{Tzitzimpasis2024}&\explbox&divergence-curl regularization &$||\nabla \text{div}(\mathbf{u})||^2+|||\nabla \text{curl}(\mathbf{u})||^2$&\dlbox\\
\hline\hline
\end{tabular}
\end{scriptsize}
\label{tab:modelbased}
\end{table}
\subsubsection{Discontinuities}
\label{sec:disc}
While many regularization techniques aim to enforce spatial smoothness, there are scenarios where allowing locally discontinuous deformations is more appropriate. Discontinuities are, for example, observed at organ boundaries \citep{nie2019}.
A popular method that models local discontinuities is total variation (TV) regularization, which originates from works on image noise removal \citep{rudin1992} and optical flow estimation of overlapping objects \citep{sun2010secrets}.
TV regularization, based on the $L_1$-norm of the deformation gradient, smoothens regions with little variation while preserving edges and discontinuities.
However, its non-differentiability at zero poses a challenge for gradient-based optimization. To address this, smooth approximations \citep{sun2010secrets} or duality-based optimization schemes \citep{pock2007} have been proposed. 

In its standard formulation, TV regularization treats each gradient direction independently, which is why it is also referred to as \textit{anisotropic} TV regularization \citep{chumchob2013, vishnevsky2017isotropic}. 
In contrast, \textit{isotropic} TV regularization \citep{zhang2016, vishnevskiy2014total, vishnevsky2017isotropic} couples the gradient directions, making it better suited for non-axis aligned motion. 
Unlike the explicit approaches above, \cite{nie2019} implicitly model local discontinuities by parameterizing the transformation with functions of bounded deformations \citep{nie2019}. 
Overall, global discontinuity-preserving regularization has limited application in conventional and learning-based registration. However, it is foundational for many spatially adaptive extensions that accommodate sliding motion (see Sec. \ref{sec:prior-knowledge-single-organ}).
\subsubsection{Physics and Biomechanics Inspired Properties}
\label{sec:phy}
Human organs conform to physical and biomechanical principles, inspiring various regularization approaches that constrain deformations to align with these properties.
Physics inspired regularization typically models the moving image $\moving{}$ as a physical object to which external forces are applied at every location. 
The resulting deformation depends on the material properties of the object. This is often modeled as a metal sheet, an elastic object, or a viscous fluid. 

The original implicit bending energy (BE) regularization in the 2D thin-plate splines registration \citep{bookstein1989} models $\moving{}$ as a thin metal sheet and minimizes the energy required for bending it based on keypoint correspondences.
Its explicit 3D equivalent has been formulated in the FFD registration \citep{rueckert1999}. 
Alternatively, $\moving{}$ can be modeled as an elastic object \citep{broit1981} or viscous fluid \citep{christensen1994} with explicit regularization. 
The linear elastic regularization introduced by \cite{broit1981} assumes a linear relationship between the strain (relative deformation) and stress (internal force). 
Recently, \cite{ringel2023} proposed regularized Kelvinlet functions as an implicit method to model linear elasticity. Kelvinlet functions are analytical solutions to the linear elastic partial differential equation (PDE), and for registration, their optimal superposition is identified. Hyperelastic regularization relaxes the linearity assumption and allows more flexibility \citep{rabbitt1995, burger2013}.

While suitable for small deformations, elasticity regularization accumulates energy with increasing external forces. 
In comparison, viscous fluid regularization \citep{christensen1994} does not accumulate internal energy, making it suitable for large deformations. 
Both elastic and fluid regularization are based on the Navier-Stokes/Navier-Cauchy PDEs, with elastic regularization operating on displacement fields $\disp{}$ and fluid regularization on velocity fields $\velocity{}$. The elasticity and viscosity properties are governed by two physical parameters $\lambda, \mu\in\mathbb{R}$ (see Tab. \ref{tab:modelbased}). 
Typically, isotropy and homogeneity of the material are assumed, meaning the material properties are independent of location and direction. 

Several learning-based registration frameworks have incorporated physics inspired regularization. 
Examples include analytical hyperelasticity equilibrium gap regularization, which penalizes deviations from equilibrium during training \citep{alvarez2024}, and using physics-informed neural networks (PINNs), which learn data-driven solutions to PDEs. 
In the context of registration, PINNs have been employed to model linear elasticity \citep{min2023} and hyper-/nonlinear elasticity \citep{ARRATIALOPEZ20235, min2024}. 
They are optimized for each image pair individually. 
Apart from that, \cite{yang2022} introduced a compressed regularization approach based on fluid mechanics. This method suppresses folding in regions of high compression,
by treating deformation grid intersections as fluid molecules. 
Recently, \cite{Tzitzimpasis2024} have proposed a divergence-curl regularization that enforces smoothness in the divergence and curl components of the deformation individually, adhering to the physical principles of fluid flow.

\subsubsection{Discussion}
Model based regularization forms the foundation of many image registration techniques. The primary goal of most methods -- including smoothness, invertibility, inverse-consistency, cycle-consistency, and diffeomorphic regularization -- is to ensure global smoothness, prevent folding, and preserve the overall topology. Yet, their successful application requires careful consideration.

Smoothness regularization can improve the robustness and speed of the registration, for instance, in multiresolution registration. Yet, excessive regularization may lead to over-smoothing, compromising alignment precision. To balance the trade-off between smoothness and registration accuracy, precise tuning of the regularization weight $\alpha$ (see Eq. \ref{eq:opt}, Fig. \ref{fig:smo}) is essential.
Invertibility constraints have been shown to effectively mitigate folding artifacts without affecting registration metrics like the Dice score \citep{estienne2021, pal2022}. While inverse-consistency is computationally demanding due to the inversion operation in every optimization iteration, cycle-consistency provides a more efficient alternative in terms of time and memory \citep{zhou2023}.

Despite the theoretical guarantees of diffeomorphic regularization, minor folding may still occur in practice due to numerical errors \citep{Avants2008, Ashburner2011, dalca2019, mok2021}. While the differentiable scaling-and-squaring layer of \citep{dalca2019} facilitates the incorporation of diffeomorphic registration into learning-based registration approaches, the computational cost of velocity field integration remains challenging. As a result, downsampling of the velocity field is often used, for instance, in \citep{dalca2019}, which can reduce the accuracy of high-resolution registration. Recent approaches, such as ResNet-based methods \citep{joshi2023r2net} or time-embedded networks \citep{matinkia2024}, avoid scaling-and-squaring altogether and offer full-resolution integration without substantial computational overhead. 

TV regularization, while effective for local discontinuities, faces the challenge of non-differentiability at zero, complicating its integration in gradient-based optimization and neural network training.
Physics inspired regularization methods offer a theoretically grounded approach; however, these methods assume isotropic and homogeneous tissue properties, which may not hold for complex biological structures. Furthermore, the selection of values for the physical parameters $\lambda,\mu$ (Tab. \ref{tab:modelbased}) is arbitrary and frequently lacks justification, limiting both interpretability and generalizability \citep{christensen1996, min2023}.
Overall, some general properties like smoothness, invertibility, and inverse-consistency are widely applicable across different anatomic structures. However, more specialized constraints, such as incompressibility or physics inspired regularization, are best suited for targeted applications where some high level information about the data is given. For example, global incompressibility can benefit the registration of contrast-enhanced images to prevent lesions of interest from shrinking during the registration process \citep{tanner2000, Rohlfing2003}. 

Many model based regularization methods have been effectively adapted to learning-based registration. The most prominent approaches include explicit regularization in the training loss function, as well as implicit smoothness regularization via multiresolution schemes and B-spline parameterization. However, methods for discontinuity preservation, incompressibility, and physical properties remain underexplored in learning-based settings despite their extensive development in conventional registration. 

In conclusion, model based regularization methods provide a powerful framework for enforcing deformation properties based on prior assumptions. However, they require careful tuning and only have limited flexibility due to their global application. 
As we will see in the following sections, model based regularization serves as the basis for problem specific and learned regularization, and maintains its essential role in the ongoing advancements of medical image registration.


\defcitealias{nlst}{NLST, 2024} 
\begin{figure}
    \centering
    \includegraphics[width=.9\textwidth]{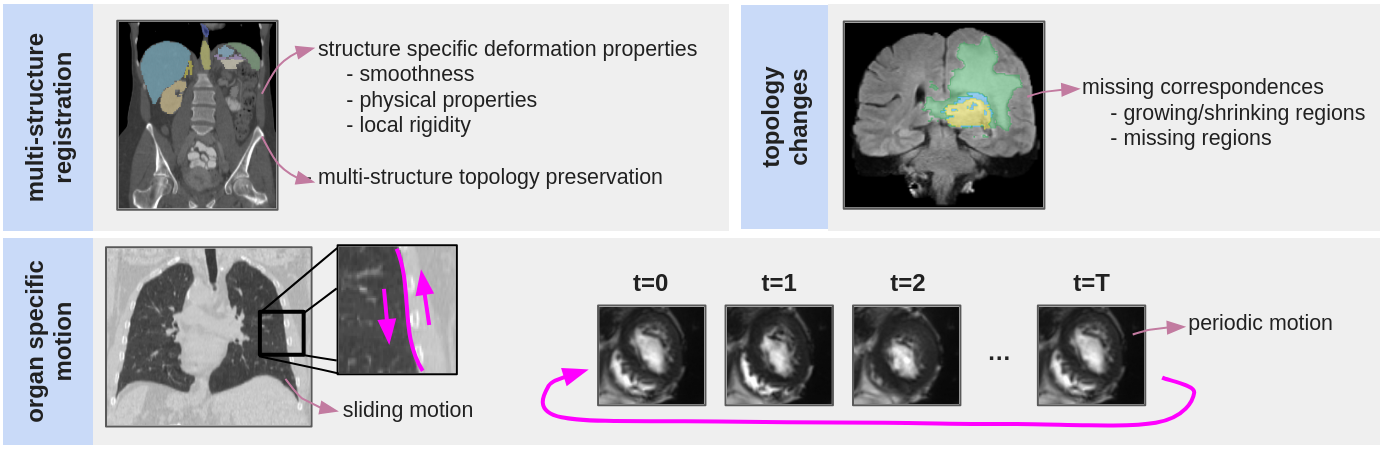}
    \caption{Problem specific regularization: Depending on the registration problem and data, different deformation properties may arise. With data-specific information, such as segmentation maps, regularization can locally account for suitable deformation properties. Images are taken from the Learn2Reg abdominal CT \citep{abdCT}, NLST \citepalias{nlst}, BraTS \citep{brats}, and ACDC \citep{acdc} datasets.}
    \label{fig:cat2}
\end{figure}

\subsection{Problem Specific Regularization with Prior Knowledge}
\label{sec:prior-knowledge}
This section presents regularization tailored to particular registration problems, organized according to the corresponding registration tasks. 
While model based regularization (Sec. \ref{sec:prior-assumptions}) imposes global deformation properties based on user \textit{assumptions}, the methods in this section enrich the registration process with data-specific \textit{knowledge}. 
Often, this includes spatial information, which makes the regularization spatially varying and adaptive to local characteristics. Additionally, knowledge about the clinical context and physiology of the anatomical structures can be used.

We divide problem specific methods into three categories of registration problems that require specific regularization: Registration of images with (i) multiple structures (Sec. \ref{sec:prior-knowledge-multistructure}), (ii) organs that exhibit special motion properties (Sec. \ref{sec:prior-knowledge-single-organ}), and (iii) changing topologies (Sec. \ref{sec:prior-knowledge-topoligical-change}). These are presented in the following and visualized in Fig. \ref{fig:cat2}.

\subsubsection{Multi-Structure Registration}
\label{sec:prior-knowledge-multistructure}
Multi-structure registration aims to accurately align multiple anatomical structures within one image (see Fig. \ref{fig:cat2}, top left). This is commonly encountered at the organ level, as in abdominal and chest registration, or at the sub-organ level, such as in brain registration. 
Two key challenges arise in these scenarios: The variability of deformation properties of individual structures and overall topology preservation of the structures, including their relative positions within the body and individual shapes. An overview of the methods discussed in this section is shown in Tab. \ref{tab:problemspecific-multistruc}.

\defcitealias{grossbroehmer2024}{Großbröhmer, 2024} 
\begin{table}
\centering
\caption{Problem specific regularization methods:  Multistructure registration. (\explbox = explicit, \implbox = implicit, \guidbox = guidance, \convbox = conventional, \dlbox = learning-based, abd. = abdominal, seg. = segmentation, reg. = registration)}
\begin{scriptsize}
\begin{tabular}{c|cp{3.6cm}cp{5.1cm}p{2.1cm}c} 
\hline\hline
\multicolumn{2}{c}{\textbf{\makecell[c]{registration\\ problem}}}  & \multicolumn{1}{c}{\textbf{\makecell[c]{reference}}} & \textbf{type}& \multicolumn{1}{c}{\textbf{\makecell[c]{approach}}} & \multicolumn{1}{c}{\textbf{\makecell[c]{modality and\\ anatomy}}}& \multicolumn{1}{c}{\textbf{\makecell[c]{registration\\ framework}}} \\
\hline
\multirow{47}{*}{\rotatebox[origin=c]{90}{\makecell[c]{multistructure registration}}} 
 & \multirow{26}{*}{\rotatebox[origin=c]{90}{\makecell[c]{structure specific \\ properties}}} & \citep{forsberg2010}& \explbox&size-adaptive Gaussian kernel& brain MR & \convbox\\
 & & \citep{wei2022}&\explbox &size-adaptive Gaussian kernel& brain MR& \dlbox\\
 &  & \citep{freiman2011}& \explbox&anisotropic diffusion& abd. CT&\convbox\\
 &  & \citep{wang2023context}& \explbox&Gaussian smoothing in folding regions& liver CT/brain MR& \dlbox\\
&  & \citep{stefanescu2004}& \explbox&varying smoothing weight& brain MR& \convbox\\
&  & \citep{ye2023}& \explbox&varying smoothing weight& cardiac MR/US& \dlbox\\
&  & \citep{wei2022}& \explbox& region-wise incompressibility & brain MR& \dlbox\\
&  & \citep{kabus2005}& \explbox&varying elasticity parameters& toy data & \convbox\\
&  & \citep{Risholm2010}& \explbox&varying elasticity parameters& brain MR& \convbox\\
&  & \citep{Drakopoulos2014}& \explbox&varying elasticity parameters& brain MR& \convbox\\
&  & \citep{Lester1999}& \explbox&varying viscosity parameters& neck MR & \convbox\\
&  & \citep{brock2005}& \implbox&varying elasticity parameters & lung/abd. MR & \convbox\\
&  & \citep{little1997}& \implbox&region-wise then fuse & neck MR & \convbox\\
&  & \citep{greene2009}& \implbox&region-wise then fuse & prostate CT& \convbox\\
&  & \citep{chen2021}& \implbox&region-wise then fuse & cardiac MR& \dlbox\\
&  & \citep{arsigny2005polyrigid}& \implbox &poly-rigid & hist. slices& \convbox\\
&  & \citep{commowick2008}& \implbox&poly-rigid/affine & brain MR/abd. CT & \convbox\\
& & \citep{edwards1998}& \implbox&varying transformations & brain MR & \convbox\\
&  & \citep{tanner2000}& \implbox&locally rigid FFD& breast MR & \convbox\\
&  & \citep{staring2007}& \explbox&successive smoothing in rigid regions & lung CT& \convbox\\
&  & \citep{loeckx2004}& \explbox&orthogonality constraint on $\jacdet$ & full-body PET & \convbox\\
&  & \citep{ruan2006}& \explbox&orthogonality constraint on $\jacdet$ & lung CT & \convbox\\
&  & \citep{staring2007-2}& \explbox&affine and orthonormality constraints & lung CT & \convbox\\
&  & \citep{modersitzki2008}& \explbox&affine and orthonormality constraints & knee CT & \convbox\\
&  & \citep{haber2009}& \explbox&reformulation of local rigidity constraints & toy data& \convbox\\
\cline{2-7}
 & \multirow{25}{*}{\rotatebox[origin=c]{90}{\makecell[c]{multistructure \\ topology preservation}}} &\citep{hu2018weakly}& \guidbox &Dice in training loss& prostate MR/US& \dlbox\\
&  & \citep{balakrishnan2019voxelmorph}& \guidbox &Dice in training loss & brain MR& \dlbox\\
&  & \citep{mok2021anatomy}& \guidbox &Dice in training loss & abd. CT/brain MR& \dlbox\\
&  & \citep{lu2023}& \guidbox &Dice in training loss & cardiac MR& \dlbox\\
&  & \citep{mahapatra2018}& \guidbox &joint reg. + seg., multitask CycleGAN & lung x-ray& \dlbox\\
&  & \citep{khor2023}& \guidbox &joint reg. + seg., cross-task attention & brain/uterus MR & \dlbox\\
&  & \citep{estienne2019}& \guidbox &joint reg. + seg., multi-decoder branches & brain MR& \dlbox\\
&  & \citep{elmahdy2021joint}& \guidbox &joint reg. + seg., multi-decoder branches & prostate CT& \dlbox\\
&  & \citep{xu2019}& \guidbox &joint reg. + seg., multi-network & brain/knee MR& \dlbox\\
&  & \citep{chen2022joint}& \guidbox &joint reg. + seg., multi-network & cardiac MR& \dlbox\\
&  & \citep{raveendran2024}& \guidbox &joint reg. + seg., multi-network & lung CT/abd. MR& \dlbox\\
&  & \citepalias{grossbroehmer2024}& \guidbox &learned segmentation encodings & abd. CT/MR& \dlbox\\
&  & \citep{rühaak2017}& \guidbox & keypoint constraints& lung CT & \convbox\\
&  & \citep{papenberg2009}& \guidbox & keypoint constraints& liver MR/CT & \convbox\\
&  & \citep{kearney2015}& \guidbox & keypoint constraints& head/neck CT & \convbox\\
&  & \citep{matkovic2024}& \guidbox & keypoint constraints& lung CT & \dlbox\\
&  & \citep{heinrich2015corrfield}& \guidbox &keypoint graph & lung CT & \convbox\\
&  & \citep{hansen2021graphregnet}& \guidbox &keypoint graph & lung CT& \dlbox\\
&  & \citep{heinrich2022voxelmorph++}& \guidbox &keypoint graph & abd./lung CT& \dlbox\\
&  & \citep{wang2023keymorph}& \guidbox &joint keypoint learning and matching & brain MR & \dlbox\\
&  & \citep{wang2024brainmorph}& \guidbox & joint keypoint learning and matching & brain MR& \dlbox\\
\hline\hline
\end{tabular}
\end{scriptsize}
\label{tab:problemspecific-multistruc}
\end{table}
\textbf{Structure Specific Deformation Properties:} To address varying deformation properties across different structures, regularization can be employed that enforces structure specific smoothness, physical properties, or rigidity.
\textit{Spatially adaptive smoothness} is concerned with locally adapting the amount of smoothness. This can be achieved with a Gaussian smoothing kernel that adapts its size and shape to local image properties, such as the eigenvalues and -vectors of the image structure tensor \citep{forsberg2010} or the uncertainty of the local displacement magnitude \citep{wei2022}.
\cite{freiman2011} derive local anisotropic smoothing from local affine transformations. Neighboring transformations with minimal differences are smoothed more since they are considered to belong to the same structure.
Similarly, locally adaptive Gaussian filters have been proposed that only target areas with folding, indicated by negative values of $\jacdet$ \citep{wang2023context}.

Apart from adapting the regularization, another possibility is to extend the scalar regularization weight $\alpha$ in Eq. \ref{eq:opt} to a location dependent weight map $\mathbf{A}(x)\in\mathbb{R}^{H\times W (\times D)}$. \cite{stefanescu2004} derive a map from local intensity variations and apply more smoothing to regions with small variations.
Learning-based registration methods have adopted spatially adaptive regularization weights, e.g. in  \cite{ye2023}, where a weight map is based on the predicted deformation during the training process. 
More smoothing is applied to image locations at which small displacements are predicted. 
Beyond smoothness, structure dependent incompressibility and anti-folding constraints have been proposed within the training loss of a registration network \citep{wei2022}. 

Moreover, it may be beneficial to model \textit{structure specific physical properties}. Physical properties, such as elasticity, not only vary across structures but also between healthy and pathological tissues. For example, increased stiffness is observed for fibrotic lung and liver tissue \citep{talwalkar2008elastography, haak2018matrix}. 
A straightforward approach to model local physical properties involves using location dependent physical parameters $\lambda(x),\mu(x)$ within physics inspired regularization (see tab. \ref{tab:modelbased}). This has been proposed for linear elastic \citep{kabus2005, Risholm2010, Drakopoulos2014} 
and viscous fluid regularization \citep{Lester1999}.
An alternative approach registers each structure individually using global model based linear elastic regularization. The obtained local deformations are then fused into a single global transformation \citep{brock2005}. 

For images containing rigid structures, \textit{local rigidity regularization} is necessary to ensure rigid deformations of rigid structures while allowing nonlinear deformations of surrounding soft tissue. This is encountered, for instance, with the skull in head-neck \citep{little1997} or the ribs in lung registration \citep{Baluwala2011}. Also, local rigidity may be desired for the registration of tumors to allow a manual comparison by an expert \citep{tanner2000, staring2007, staring2007-2}.
An implicit approach to address local rigidity is the spatially varying parameterization of the deformation model.
For example, in \cite{little1997} and \cite{greene2009} the structures are individually registered with either rigid/affine or nonlinear deformations depending on their characteristics before being fused into a global deformation. To assist the fusion process, nonlinear deformations can be additionally constrained to vanish near the boundaries of rigid structures \citep{greene2009}. 
Further extensions include fuzzy rigid regions that are parameterized by Gaussians centered around anchor points \citep{arsigny2005polyrigid} and ensuring global invertibility of the fused deformation in log-Euclidean polyaffine approaches \citep{commowick2008, arsigny2009fast}. 
Locally rigid FFD registration is obtained with control point coupling within each rigid region in \cite{tanner2000}. 

Explicit regularization methods include the use of spatially varying weight maps that locally switch between different types of transformations, (rigid, fluid, and soft tissue deformations) \citep{edwards1998}, and the successive application of a smoothing filter that averages the displacements and converges to a constant displacement within each rigid region \citep{staring2007}. 
The latter approach has been combined with linear elasticity regularization in \cite{Baluwala2011}. 

To promote local rigidity, explicit variational loss terms can be further employed based on the Jacobian determinant $\jacdet$.
A deformation is locally rigid if $\jacdet$ is orthogonal, i.e., $(\jacdet)^T\jacdet=1$. Consequently, local deviations of $\jacdet$ from orthogonality are be penalized in rigid regions in \cite{loeckx2004, ruan2006}. Building upon this, the constraints proposed by \cite{staring2007-2, modersitzki2008} additionally enforce local affine-ness and orientation-preservation by penalizing deviations of the second-order derivatives $\partial^2\disp$ from zero, as well as deviations of $\jacdet$ from unity. This combination of constraints shows better local rigidity preservation than their individual application.
\cite{haber2009} have proposed a further method that guarantees local rigidity by reformulating locally rigid registration as an unconstrained problem. It parameterizes rigid deformations directly at pixel locations of rigid structures.

Despite extensive exploration in conventional registration frameworks, local rigidity is less commonly adapted to learning-based settings. A notable exception is the work of \cite{chen2021}, where a multichannel registration network predicts structure specific velocity fields that are fused together. Since no smoothing is applied during fusion, local discontinuities are preserved at the organ boundaries. Structure specific physical properties remains equally unexplored.

\textbf{Multi-Structure Topology and Shape Preservation:}
Under natural body motion, the relative positions and shapes of the individual structures should remain consistent. Consequently, preserving the overall inter-structure topology and shapes is critical in multistructure registration.
Segmentation maps are particularly useful for guiding registration tasks since they provide detailed location and shape information. By directly incorporating segmentation maps in the optimization process, robustness to variations in image intensity and complex deformations can be increased. 

The state-of-the-art approach to incorporating segmentation maps for image registration uses overlap measures as guiding loss terms in the objective or training function. Typically, the Dice score is used, as is frequently found in learning-based registration, e.g., in \cite{hu2018weakly, hu2018isbi, balakrishnan2019voxelmorph, mok2021anatomy, lu2023}. Since ground truth segmentations are required during training, this is called weak supervision.
To alleviate this requirement, other works adopt multitask strategies to jointly learn the segmentation and registration from input images. 
This approach has been implemented across various network architectures, including CycleGANs \citep{mahapatra2018}, cross-task attention mechanisms \citep{khor2023}, encoder-decoder architectures with dual decoder branches \citep{estienne2019, elmahdy2021joint} and paired networks that interact during training \citep{xu2019, chen2022joint, raveendran2024}.
Notably, \cite{khor2023} propose a network that locally adapts the influence of segmentation guidance on the registration based on the segmentation overlap, enhancing registration performance.

Recently, capsule networks \citep{sabour2017} have been leveraged within image registration to model part-whole relationships among anatomical structures \citep{yan2024}. These networks capture detailed pose and shape information, making them well suited for incorporating relative relationships between structures and their components. Finally, \cite{grossbroehmer2024} introduce a registration network that is based on learned segmentation encodings rather than image features, demonstrating high generalizability.

Beyond segmentation maps, corresponding keypoint distance can also serve as a guiding mechanism for registration. Keypoints can enhance registration robustness by providing reliable anchors across and within structures. 
Although this approach is less popular, it is found, for instance, in \cite{rühaak2017, papenberg2009, kearney2015}. 
Moreover, graphs can be constructed from keypoints to guide a sparse registration \citep{heinrich2015corrfield, hansen2021graphregnet, heinrich2022voxelmorph++}. From the sparse result, a dense deformation is then derived, for instance, with thin plate splines. 
Recently, \cite{wang2023keymorph, wang2024brainmorph} have introduced registration networks capable of jointly detecting and matching keypoints for registration guidance, and \cite{matkovic2024} have proposed to incorporate keypoint guidance in a GAN architecture. 

\begin{table}[t]
\centering
\caption{Problem specific regularization methods: organ specific registration. (\explbox = explicit, \implbox = implicit, \convbox = conventional, \dlbox = learning-based, abd. = abdominal, Id. = identity)}
\begin{scriptsize}
\begin{tabular}{c|cp{3.8cm}cp{5.1cm}p{2cm}c} %
\hline\hline
\multicolumn{2}{c}{\textbf{\makecell[c]{registration\\ problem}}}  & \multicolumn{1}{c}{\textbf{\makecell[c]{reference}}} & \textbf{type}& \multicolumn{1}{c}{\textbf{\makecell[c]{approach}}} & \multicolumn{1}{c}{\textbf{\makecell[c]{modality and\\ anatomy}}}& \multicolumn{1}{c}{\textbf{\makecell[c]{registration\\ framework}}} \\
\hline
\multirow{44}{*}{\rotatebox[origin=c]{90}{\makecell[c]{organ specific registration}}} & \multirow{34}{*}{\rotatebox[origin=c]{90}{\makecell[c]{sliding motion}}} & \citep{yin2010}& \explbox&suppress smoothing near boundary & lung CT& \convbox\\
&  & \citep{jud2017directional}& \explbox&suppress smoothing near boundary & lung CT& \convbox\\
&  & \citep{heinrich2010}& \explbox&$L_p$-norm& lung CT& \convbox\\
&  & \citep{Gong2020}& \explbox&$L_p$-norm& lung CT/PET& \convbox\\
&  & \citep{duan2023}& \explbox&$L_p$-norm& lung CT& \dlbox\\
&  & \citep{luo2023}& \explbox&$L_p$-norm& lung CT& \dlbox\\
&  & \citep{ruan2008}& \explbox&anisotropic diffusion& lung CT& \convbox\\
&  & \citep{schmidtrichberg2009}& \explbox&anisotropic diffusion& lung CT& \convbox\\
&  & \citep{pace2011}& \explbox&anisotropic diffusion& lung CT& \convbox\\
&  & \citep{pace2013}& \explbox&anisotropic diffusion& lung CT& \convbox\\
&  & \citep{tanner2013}& \explbox&anisotropic diffusion& abd. MR& \convbox\\
&  & \citep{delmon2013}& \explbox&anisotropic diffusion& lung CT& \convbox\\
&  & \citep{risser2013}& \explbox&anisotropic diffusion& lung CT& \convbox\\
&  & \citep{jud2016bi}& \explbox&anisotropic diffusion& lung CT& \convbox\\
&  & \citep{schmidrichberg2012}& \explbox&anisotropic diffusion& lung CT& \convbox\\
&  & \citep{risser2011}& \implbox&region-wise in tangential direction & lung CT& \convbox\\
&  & \citep{Papiez2014}& \explbox&bilateral filter& lung CT& \convbox\\&& \citep{papiezSLIC}& \explbox&adaptive Gaussian filter& liver CT& \convbox\\ 
&  & \citep{papiez2018gifted}& \explbox&adaptive Gaussian filter& lung CT/liver MR& \convbox\\  
&  & \citep{berendsen2014sliding}& \implbox&region-wise then fuse, boundary constraint & lung CT& \convbox\\
&  & \citep{derksen2015}& \implbox&region-wise then fuse, boundary constraint & lung CT& \convbox\\
&  & \citep{Preston2016}& \implbox&region-wise then fuse, label constraint& lung CT& \convbox\\
&  & \citep{eiben}& \implbox&region-wise then fuse, boundary constraint& lung CT& \convbox\\
& & \citep{ruan2009}& \explbox&constraints on Helmholtz-Hodge components& lung CT& \convbox\\
&  & \citep{ai2019}& \explbox&constraints on Helmholtz-Hodge components& abd. CT& \convbox\\
&  & \citep{sandkuehler2018}& \explbox&graph diffusion, zero cross-boundary weights& lung CT& \convbox\\
&  & \citep{jud2018}& \explbox&graph TV& lung CT& \convbox\\
&  & \citep{andrade2021}& \explbox&elasto-plasticity& lung CT& \convbox\\
&  & \citep{ng2020}& \explbox&penalize non-parallel motion& lung CT& \dlbox\\
&  & \citep{lu2023}& \explbox&suppress smoothing near boundaries& cardiac MR& \dlbox\\
&  & \citep{duan2023}& \explbox&$L_p$-norm, boundaries are learned& cardiac MR& \dlbox\\
\cline{2-7}
 & \multirow{11}{*}{\rotatebox[origin=c]{90}{\makecell[c]{cyclic motion}}} & \citep{shen2005}& \explbox&smoothness between last/first image & cardiac MR& \convbox\\
&  & \citep{ledesma2005} & \explbox&smoothness between last/first image & cardiac US & \convbox\\
&  & \citep{metz2011} & \explbox&smoothness between last/first image & lung/cardiac CT & \convbox\\
&  & \citep{vandemeulebroucke2011} & \explbox&smoothness between last/first image & lung CT & \convbox\\
&  & \citep{ye2023} & \explbox&smoothness between last/first image & cardiac US/MR& \dlbox\\
&  & \citep{brehm2012} & \explbox&penalize deformation cycle $\neq$ Id. & lung CT& \convbox\\
&  & \citep{fechter2019} & \explbox&penalize deformation cycle $\neq$ Id. & lung CT& \convbox\\
&  & \citep{bai2009} & \explbox&cyclic B-Spline coefficients & lung PET& \convbox\\
&  & \citep{mceachen2000} & \implbox&cyclic basis functions in FFD & cardiac MR& \convbox\\
&  & \citep{wiputra2020} & \implbox&cyclic basis functions in FFD & cardiac US& \convbox\\
\hline\hline
\end{tabular}
\end{scriptsize}
\label{tab:problemspecific-organ}
\end{table}

\subsubsection{Organ Specific Registration with Special Motion Properties}
\label{sec:prior-knowledge-single-organ}
While multistructure registration often considers different organs at the same time, other scenarios focus on single organs only. For instance, intra-patient cardiac and lung registration is employed to analyze motion dynamics from spatio-temporal data, with applications in functional analysis or radiotherapy treatment planning \citep{keall2005four, reinhardt2008,choi2013lung}. 
However, these organs exhibit unique deformation characteristics that are critical to consider during registration.
Specifically, cardiac and respiratory motion is periodic and quasi-cyclic in nature and  breathing induces sliding motion at organ boundaries (Fig. \ref{fig:cat2}, bottom). An overview of the methods discussed in this section is shown in Tab. \ref{tab:problemspecific-organ}.

\textbf{Sliding motion:}
At organ boundaries, motion in opposing directions leads to local discontinuities, such as the sliding of lungs, diaphragm, and liver against the pleural wall or between lung lobes along fissures. This cannot be modeled with standard regularization.
Methods that address sliding motion combine smoothness and discontinuity regularization and either adapt to (a) boundary proximity, (b) deformation direction, or (c) use region-wise registration. 

Explicit diffusion regularization can be adapted to allow discontinuities near sliding boundaries while maintaining smoothness elsewhere. 
This is achieved by \textit{suppressing smoothing} near boundaries \citep{yin2010, jud2017directional} or by switching from diffusion ($L_2$-norm) to TV ($L_1$-norm) regularization. This is called $L_p$-norm regularization ($1<p<2$) and encountered in conventional \citep{heinrich2010, Gong2020} and learning-based registration \citep{duan2023, luo2023}.

Apart from boundary distance, the \textit{deformation direction} can be considered: Sliding motion requires tangential smoothness along boundaries while exhibiting discontinuities in the normal direction. 
A common method to address this is to leverage anisotropic diffusion that decomposes smoothing into tangential and normal components: Isotropic smoothing is applied within organs and anisotropic smoothing that vanishes in the normal direction is applied near boundaries \citep{ruan2008, schmidtrichberg2009, pace2011, pace2013, tanner2013, delmon2013, risser2013, jud2016bi}. 
Anisotropic smoothing is extended in \cite{schmidrichberg2012} to locally adapt to the estimated amount of slippage. To this end, the deformation variations are compared at sampled points on both sides of the boundary. 
Another approach is to employ global smoothing in the normal direction while performing tangential smoothing independently at each side of the boundary \citep{risser2011}.
Anisotropic smoothing can furthermore be achieved with the bilateral filter proposed in \cite{Papiez2014}. This filter balances local discontinuities and smoothness by adapting to intensity differences, spatial smoothness, and local deformation field similarities. 
In \cite{fu2018} the bilateral filter is applied in the tangential direction only, while isotropic normal smoothing is used. 
Adaptive Gaussian filters derived from super-voxel clusters of the image further enhance sliding boundary fidelity by reducing smoothing in areas of high intensity or deformation variance \citep{papiezSLIC, papiez2018gifted}. Since supervoxel clusters capture local discontinuities, the filters can effectively transfer discontinuity information to the deformation.

Besides this, sliding motion can be addressed with \textit{region-wise} registration, such as in 
\citep{risser2013,berendsen2014sliding, derksen2015, Preston2016, eiben}. 
First, smooth registration is performed independently for each side of a sliding boundary. 
Then, the resulting deformations are fused into a global deformation. To preserve discontinuities at boundaries, no smoothing is used during fusion. 
A major challenge with this approach is the presence of gaps and overlaps at the sliding boundaries. Solutions include penalizing misalignment of boundaries during the region-wise registration \citep{berendsen2014sliding, derksen2015, eiben} or constraining the deformed segmentation maps to a single label per voxel \citep{Preston2016}.

Further advancements in sliding motion regularization include component-wise regularization of the divergence-free, curl-free, and harmonic components of the deformation (which can be obtained with the Helmholtz decomposition) to preserve large shears along boundaries \citep{ruan2009, ai2019}.
Graph-based methods extend diffusion regularization by assigning zero-weight edges across sliding boundaries \citep{sandkuehler2018} or adapt TV regularization \citep{jud2018}.
Physics inspired regularization has been leveraged for sliding motion in \cite{andrade2021} by combining linear elastic and van Mises plasticity models in the regularization to allow high shear deformations along sliding boundaries. 
Region-wise thin-plate spline and B-spline registration have been proposed in \cite{xie2011,hua2017}.

Although extensively studied in conventional registration, sliding motion regularization has seen limited adoption in learning-based registration frameworks. Explicit sliding motion regularization has been proposed for network training, including penalization of non-parallel motion of neighboring voxels \citep{ng2020}, smoothing suppression at sliding boundaries with a binary mask \citep{lu2023} and $L_p$-norm regularization \citep{duan2023}. Additionally, multitask learning is leveraged to learn boundary locations jointly with the registration in \cite{duan2023}.

\textbf{Cyclic motion}:
Registration of 4D (3D+t) image data often involves capturing a patient's cardiac or respiratory cycles, for instance, in the context of cardiac function analysis, 
\citep{cardiacanalysis}, radiation therapy
\citep{radiotherapytracking} and motion correction \citep{spieker2024}.
For such data, incorporating the cyclic nature of the deformation in the registration is particularly important.
Periodicity can be explicitly accounted for by ensuring that the last image in a sequence maps back to the first image seamlessly, as found in conventional \citep{shen2005, ledesma2005, metz2011, vandemeulebroucke2011} and learning-based \citep{ye2023} registration.
Alternatively, deviations from the full-cycle deformation to the identity deformation can be penalized \citep{brehm2012, fechter2019}.

Implicit cyclic extensions to FFD registration use periodic B-spline coefficients \citep{bai2009} and cyclic basis functions such as harmonic sinusoid or Fourier functions \citep{mceachen2000, wiputra2020}. 
It is observed that, in general, despite the prevalence of 3D cardiac and lung image registration studies, relatively few are concerned with cyclic regularization of 3D+t data. Cyclic regularization in the context of learning-based registration remains to be explored.

\subsubsection{Registration of Images with Topological Change}
\label{sec:prior-knowledge-topoligical-change}
A common assumption in image registration is that the underlying topology of the image remains unchanged, implying that every location in $\moving{}$ corresponds to a location in $\fixed{}$. 
However, there are scenarios where this assumption is not true,
particularly in clinical settings where topology changes can arise for various reasons. 
Examples include changing tissue due to lesion development \citep{niethammer2011,anton2022}, missing regions after surgical resection \citep{risholm2009, nithiananthan2012,chen2015,wodzinski2021},
variable gastrointestinal contents \citep{suh2011}, and removal of medical devices \citep{berendsen2013}.
In such cases, including prior knowledge about the clinical context in the registration is valuable, particularly when combined with spatial information about topologically changing regions.
An overview of the methods discussed in this section is shown in Tab. \ref{tab:problemspecific-topo}.

 \begin{table}[t]
\centering
\caption{Problem specific regularization methods: Registration of images with topological changes. (\explbox = explicit, \implbox = implicit, \convbox = conventional, \dlbox = learning-based, abd. = abdominal)}
\begin{scriptsize}
\begin{tabular}{c|cp{3.8cm}cp{5cm}p{2.1cm}c} %
\hline\hline
\multicolumn{2}{c}{\textbf{\makecell[c]{registration\\ problem}}}  & \multicolumn{1}{c}{\textbf{\makecell[c]{reference}}} & \textbf{type}& \multicolumn{1}{c}{\textbf{\makecell[c]{approach}}} & \multicolumn{1}{c}{\textbf{\makecell[c]{modality and\\ anatomy}}}& \multicolumn{1}{c}{\textbf{\makecell[c]{registration\\ framework}}} \\
\hline
\multirow{20}{*}{\rotatebox[origin=c]{90}{\makecell[c]{topology changes}}} 
 & \multirow{9}{*}{\rotatebox[origin=c]{90}{\makecell[c]{growing/shrinking\\ regions}}} & \citep{brett2001}& \explbox&cost function masking & brain MR& \convbox\\
& & \citep{niethammer2011}& \implbox&geometric metamorphosis & brain MR& \convbox\\
& &\citep{anton2022} & \explbox&region-limited metamorphosis&brain MR & \convbox\\
& &\citep{wang2023meta} & \explbox&region-limited metamorphosis & brain MR & \dlbox\\
& & \citep{bone2020}& \implbox&metamorphic autoencoder & brain MR& \dlbox\\
& & \citep{maillard2022}& \implbox&ResNet-based metamorphosis &brain MR& \dlbox \\
& & \citep{joshi2023r2net}& \implbox&ResNet-based metamorphosis &brain MR/liver CT& \dlbox \\
& & \citep{dong2023}& \explbox&constrain lesion volume change &brain MR/abd. CT & \dlbox \\
\cline{2-7}
& \multirow{11}{*}{\rotatebox[origin=c]{90}{\makecell[c]{missing \\ regions}}} & \citep{berendsen2014}& \explbox&minimize volume& cervical MR& \convbox\\
& & \citep{chen2015}& \explbox&minimize volume& brain MR& \convbox\\
& & \citep{wodzinski2021}& \explbox&minimize volume& breast MR& \dlbox\\
& & \citep{risholm2009}& \explbox&diffusion sink & brain MR& \convbox\\
& & \citep{suh2011}& \implbox&auxiliary deformation dimension & colorectal CT& \convbox\\
& & \citep{nithiananthan2012}& \implbox&auxiliary deformation dimension & head CT& \convbox\\ 
& & \citep{Nielsen2019}& \implbox&expand voids from pre-defined points & brain MR & \convbox\\
& & \citep{czolbe2021}& \explbox&cost function masking in training & brain MR& \dlbox\\
& & \citep{mok2022corr}& \explbox&IC loss-based cost function masking & brain MR& \dlbox\\
& & \citep{wodzinski2023}& \explbox&IC loss-based cost function masking & brain MR & \dlbox\\
& & \citep{feng2024}& \explbox& IC loss-based corrected attention layers& brain MR& \dlbox\\
\hline\hline
\end{tabular}
\end{scriptsize}
\label{tab:problemspecific-topo}
\end{table}

\textbf{Growing and Shrinking Regions:}
If an area with missing correspondences is partially present in $\moving$ and $\fixed$, this implies region growth or shrinkage. Typically this involves smooth tissue change due to the development of pathologies.
One approach that addresses such topology changes is cost function masking that restricts the optimization to topologically consistent regions \citep{brett2001}. This ensures that changing regions do not bias the registration.

Alternatively, local topology changes are accounted for through adaptations of the metamorphosis framework, which disentangles deformations from appearance changes in the loss function \citep{trouve2005, younes2010}.
For instance, geometric metamorphosis \citep{niethammer2011} models topology changes with a geometric model decoupled from the deformation of the surrounding tissue. This isolation enables the quantification of topology change which is valuable for downstream applications such as automatic lesion growth analysis. In \cite{anton2022}, the framework is refined to limit geometric changes to predefined regions. 

Metamorphic registration has also been adapted to learning-based registration. The metamorphic autoencoder of \cite{bone2020} decouples low-dimensional representations of shape and appearance changes in latent space, and metamorphic ResNet architectures are found in \cite{maillard2022, joshi2023r2net}.
Although these methods require prior segmentation maps of topologically changing regions, the metamorphic registration network by \cite{wang2023meta} eliminates this dependency by jointly learning the location of appearance changing regions.
Further approaches combine registration with predicting a quasi-normal (pathology-free) image, enabling registration between pathological and healthy images \citep{han2020}, and explicit regularization that ensures that the volume change of lesions matches the one of surrounding tissue \citep{dong2023}.

\textbf{Missing Regions:}
\label{sec:prior-knowledge-missing-regions}
In contrast to growing regions, missing regions are entirely absent in one of the two images. Such a scenario is found, for instance, in the registration of pre-to-post resection images. 
To address this, explicit regularization can be employed to minimize the volume of regions without correspondence in the other image \citep{berendsen2014,chen2015, wodzinski2018, wodzinski2021}. 
Another option is to use a diffusion sink that restricts diffusion regularization such that image forces originating within a resected region diffuse outward to surrounding areas but not in the reverse direction \citep{risholm2009}.

Implicit regularization methods that leverage the parameterization of the transformation model include modeling tissue removal in an auxiliary deformation dimension \citep{suh2011,nithiananthan2012} or by expanding voids from selected image locations \citep{nielsen2019meta}. 
To alleviate the need for prior segmentation maps, joint registration and segmentation of missing regions has been further proposed for conventional \citep{periaswamy2006, chitphakdithai2010, chen2015} and learning-based registration with conditional autoencoders \citep{czolbe2021} and convolutional neural networks (CNNs) \citep{mok2022corr, wodzinski2023, liu2023}. Here, cost function masking is applied to suppress the influence of missing regions on the registration. 

An elegant recently proposed approach automatically identifies missing regions with model based regularization within the training of registration networks \citep{mok2022corr, wodzinski2023}: Local areas with a high inverse-consistency regularization loss (Sec. \ref{sec:ic}) are interpreted as missing in the other image. This information is then used for masking or weighting the training loss. Another weighting-based method is used in the bi-directional registration network by \cite{feng2024} where a weight map guides the attention layers to focus on pathological and missing regions.

\subsubsection{Discussion}
Problem specific regularization extends beyond the global, general-purpose constraints of model based approaches by integrating data knowledge and adapting to local deformation characteristics. The spatial adaptivity enables more realistic modeling of complex deformations
which global regularization methods fail to capture.
The majority of the problem specific methods extend model based regularization methods presented in Sec. \ref{sec:prior-assumptions} to spatial adaptivity, either by location dependent regularization weight maps, as seen in sliding motion regularization, or location dependent formulation of the regularization itself, for instance, to accomplish local rigidity. 

Despite their clear advantages, problem specific regularization methods face challenges that should be considered for successful application. 
A major challenge is that many of the presented methods rely on accurate prior segmentation maps or boundary definitions, for instance,  \cite{risser2013, schmidrichberg2012}. While manual annotations by clinicians provide high precision, they are costly to obtain. This is particularly true for methods embedded within learning-based registration frameworks, where segmentation maps of the full training dataset are often required for training. 
To mitigate this, segmentations can be inferred directly from images or estimated deformations, or multitask learning can be employed to jointly learn the registration and the spatial information, such as in \cite{czolbe2021,wang2023meta, duan2023}. Moreover, the rapid advancements of open-source, pre-trained segmentation networks like TotalSegmentator \citep{totalsegmentator} offer a convenient solution for automatic segmentations that can be used in registration frameworks without notable computation overhead. 
However, such segmentations are probably not as precise as manual ones. Consequently, to reduce the dependency of registration results on annotation quality, regularization methods need to be adapted to noisy segmentations, as in
\cite{Gong2020, raveendran2024}. 

Apart from that, problem specific methods are tailored to address specific anatomical or pathological characteristics. On the one hand, careful parameter tuning might be required, for instance for local physical parameters in \cite{andrade2021}, hyperparameters that balance geometric change and deformations in \citep{bone2020}, or hyperparameters of Gaussian filters in \citep{Papiez2014}. On the other hand, different methods have different requirements about the specific scenario. For example, when addressing topology changes, some methods can only register healthy to pathological images, for instance, \cite{francois2021}, while others can only register in the other direction, e.g., \cite{maillard2022, han2020}. Consequently, their generalizability to new applications and data that slightly differ from the validation scenarios is limited. In particular, the high dependency on parameter tuning complicates registration network training, making their clinical application challenging. 

Overall, a gap in regularization method transfer from conventional to learning-based registration is observed. For multistructure registration, local physical properties and rigidity have seen limited integration into learning-based registration. In contrast, segmentation guidance can be considered state-of-the-art in learning-based registration. Sliding motion, though well-studied in conventional registration, also remains to be explored more with registration networks. 
In particular, integrating local physical properties and rigidity could drive the capabilities of registration networks forward since many applications face rigid structures or varying material properties. 

In summary, the capability of problem specific regularization to address local deformation characteristics and real-world clinical scenarios makes it indispensable for targeted applications.
However, methods heavily rely on design choices and hyperparameter tuning.
This highlights the need for more flexible regularization frameworks capable of automatically identifying deformation properties from the data itself, which is addressed with the learned regularization methods in the following section.


\subsection{Learned Regularization}
\label{sec:learned-knowledge}
This section presents and discusses the rapidly growing field of learned regularization, organized by what is learned from training data. 
While problem specific regularization incorporates data knowledge and adapts to local deformation characteristics, it still requires careful user design and parameter tuning, which can limit scalability and adaptability. 
In contrast, \textit{learned} regularization leverages data-driven techniques to infer local properties directly from training data.
For this, the regularization component of a registration framework is parameterized with a machine or deep learning model (see, for instance, Fig. \ref{fig:hypernet}).
We identify three subcategories of learned regularization: Methods that learn (i) local smoothness properties (Sec. \ref{sec:learned-smoothing}), (ii) feasible deformation spaces (Sec. \ref{sec:learned-spaces}), and (iii) test time adaptive regularization (Sec. \ref{sec:learned-weights}). An overview of learned methods is given in Tab. \ref{tab:learned}.

\subsubsection{Learned Local Smoothing and Discontinuities}
\label{sec:learned-smoothing}
Spatially varying smoothing, which automatically adapts to local image properties, can be effectively learned using deep neural networks. 
For instance, \cite{niethammer2019} have proposed a shallow CNN within a conventional diffeomorphic LDDMM framework (see Sec. \ref{sec:diff}). 
The CNN takes a momentum vector field and an image as inputs. It predicts a smoothed vector field by learning the weights of a spatially adaptive multi-Gaussian smoothing kernel. 
To ensure that the regularized output remains diffeomorphic and transfers edge information from the image to the deformation, the learned weights are additionally prevented from degenerating through total variation (TV) meta-regularization. By bounding the variance of the weights, a desired smoothing level can be further specified.
Optimization is performed in two stages: First, only the momenta are optimized while the Gaussian weights are carefully chosen, then the CNN parameters and the initial momentum vector fields are jointly optimized. An extension of this approach to spatio-temporal velocity fields has been proposed in \cite{shen2019region}. Here, two separate CNNs predict the initial momentum vector field and the smoothing weights, respectively.

A more generalized approach is presented in \cite{safadi2021}, where the regularization is parameterized as a trainable convolutional layer applied to the predicted deformation of a registration network. This method aims to minimize folding and increase anatomical plausibility.
Building upon the Reproducing Kernel Hilbert Space (RKHS) theory, the authors demonstrated that constraining the convolutional filters to be positive, semi-definite, and radially symmetric approximates RKHS regularization. These constraints are explicitly enforced in the meta-regularization of the regularization layers.
By imposing these constraints on the learned filters, they offer learned spatially adaptive smoothing filters that are highly generalizable and can adapt well to large deformations.

While the methods presented above focus on locally adapting the smoothing strength, local discontinuities are learned in \cite{jia2022, lai2023}. Using a variable splitting scheme and an auxiliary splitting variable, the registration problem of Eq. \ref{eq:opt} is divided into two simpler subproblems: a similarity and a regularization component. These subproblems are individually parameterized by a neural network, where a ResNet-based denoising network is employed for the regularization part. Optimization is performed by alternating parameter updates of the two networks or joint training. Conveniently, regularization hyperparameters such as $\alpha$ and additional weights resulting from the variable splitting scheme can be absorbed in the denoising network, alleviating manual tuning.
The regularization network effectively learns local TV regularization that avoids over-smoothing and allows for local discontinuities \citep{jia2022} or focuses on modeling sliding motion \citep{lai2023}.

\defcitealias{shuaibu2024}{Shuaibu et al., 2024} 
\begin{table}[t]
\centering
\caption{Learned regularization methods. (\explbox = explicit, \implbox = implicit, \guidbox = guidance, \convbox = conventional, \dlbox = learning-based, abd. = abdominal, phys. params. = physical parameters)}
\begin{scriptsize}
\begin{tabular}{c|p{3cm}cllp{2.5cm}c} 
\hline\hline
\multicolumn{1}{c}{\textbf{\makecell[c]{learns}}}  & \textbf{reference} & \textbf{type}& \multicolumn{1}{c}{\textbf{\makecell[c]{regularization \\ model}}}& \textbf{approach} & \multicolumn{1}{c}{\textbf{\makecell[c]{modality and\\ anatomy}}}& \multicolumn{1}{c}{\textbf{\makecell[c]{registration \\ framework}}} \\
\hline
\multirow{5}{*}{\rotatebox[origin=c]{90}{\makecell[c]{local \\smo. /disc.}}} 
 & \citep{niethammer2019}&\explbox& CNN & local multi-Gaussian weights& brain MR&\convbox \\
& \citep{shen2019region}&\explbox& two CNNs&  local multi-Gaussian weights & knee MR/lung CT & \convbox\\
& \citep{safadi2021}&\explbox& convolution layer& RKHS convolution filter & cardiac US/lung x-ray& \dlbox\\
 & \citep{jia2022}&\explbox& denoising ResNet& variable splitting & cardiac MR& \dlbox\\
& \citep{lai2023}&\explbox& denoising ResNet&variable splitting  & brain MR/lung CT& \dlbox\\
\hline
\multirow{30}{*}{\rotatebox[origin=c]{90}{\makecell[c]{feasible\\ deformation space}}} 
   & \citep{wouters2006}&\implbox& PCA& opt. w.r.t. PCA coeffs.& brain MR& \convbox\\
 & \citep{hu2008}&\implbox& PCA& opt. w.r.t. PCA coeffs.& prostate US&\convbox\\
 & \citep{qiu2012}&\implbox& PCA& opt. w.r.t. PCA coeffs.& brain MR& \convbox\\
 & \citep{zeng2018}&\implbox& PCA& opt. w.r.t. PCA coeffs.& brain MR& \convbox\\
 & \citep{loeckx2003}&\implbox& PCA& PCA on FFD control points& lung CT& \convbox\\
 & \citep{kim2008}&\guidbox& PCA& PCA for initialization & brain MR& \convbox\\
 & \citep{tanner2009}&\guidbox& PCA& PCA for initialization & breast MR/x-ray& \convbox\\
 & \citep{cui2017}&\explbox& PCA& constrained PCA coeffs.& lung SPECT& \convbox\\
 & \citep{albrecht2008}&\explbox& PCA& constrained PCA coeffs.& upper leg x-ray&\convbox \\
 & \citep{berendsen2013}&\explbox& PCA& constrained PCA coeffs.& cervical MR& \convbox\\
 & \citep{gu2021}&\explbox& PCA& constrained PCA coeffs.& brain MR& \dlbox\\
 & \citep{hu2015}&\explbox& PCA& 4D motion model& rectal US& \convbox\\
  & \citep{jud2017}&\implbox&reproducing kernel&4D motion model & lung CT& \convbox\\
& \citep{onofrey2015}&\implbox& PCA& 4D FFD model& prostate&\convbox \\
   & \citep{xue2006}&\explbox& two PCAs& $\jacdet$ + $\disp$ PCA& brain MR & \convbox\\
   & \citep{mohamed2006}&\explbox& two PCAs& healthy+pathological PCA& brain MR& \convbox\\
 & \citep{tang2018}&\implbox& local PCAs& patch-based PCAs& brain MR& \convbox\\
& \citep{gao2021}&\implbox& autoencoder& learned PCA coeffs.& upper leg CT& \convbox\\
& \citep{hu2018}&\explbox& GAN& FEM-based adversarial loss& prostate MR/US& \dlbox\\
& \citep{bhalodia2019}&\explbox& autoencoder& reconstruction loss& brain MR& \dlbox\\
& \citep{mansilla2020}&\explbox& autoencoder& reconstruction loss& lung x-ray& \dlbox\\
& \citep{sang2020isbi}&\explbox&autoencoder & reconstruction loss& cardiac MR& \dlbox\\
& \citep{qin2020}&\explbox& autoencoder& reconstruction loss& cardiac MR& \dlbox\\
& \citep{sang2021}&\explbox& autoencoder&reconstruction loss& lung CBCT& \dlbox\\
& \citep{sang2022}&\explbox&autoencoder&  reconstruction loss& cardiac CTA/MR& \dlbox\\
& \citep{sang2020}&\implbox&decoder-only & opt. w.r.t latent vector& cardiac MR/lung CT& \convbox\\
& \citep{qin2023}&\implbox& decoder-only& opt. w.r.t latent vector& cardiac MR& \convbox\\
\hline
\multirow{9}{*}{\rotatebox[origin=c]{90}{\makecell[c]{test time \\regularization}}} 
& \citep{mok2021conditional}&\explbox& CIN layers&learn effect of $\alpha$ (diffusion)& brain MR& \dlbox\\
& \citep{hypermorph}&\explbox& hypernet& learn effect of $\alpha$ (diffusion)&brain MR & \dlbox\\
& \citep{wang2023context}&\explbox& CIN layers&learn effect of $\alpha$ (diffusion)&brain MR &\dlbox \\
& \citep{chen2023spatially}&\explbox& CIN/hypernet&learn effect of $\alpha$ (diffusion) & brain MR& \dlbox\\
& \citep{zhu2024hierarchical}&\explbox& CIN layers&learn effect of $\alpha$ (diffusion)& brain MR/lung CT& \dlbox\\
& \citepalias{shuaibu2024}&\explbox& MLP&learn optimal $\alpha$(diffusion) &brain MR &\dlbox\\
& \citep{reithmeir2024}&\explbox& hypernet&learn effect of phys. params. & lung CT/cardiac MR& \dlbox\\
& \citep{reithmeir2024spie}&\explbox& hypernet&learn effect of phys. params. & lung CT& \dlbox\\
& \citep{xu2022miccai}&\explbox& mean-teacher net& dynamic $\alpha$ during training& abd. MR/CT& \dlbox\\
\hline\hline
\end{tabular}
\end{scriptsize}
\label{tab:learned}
\end{table}

\begin{figure}
    \centering
    \includegraphics[width=.9\textwidth]{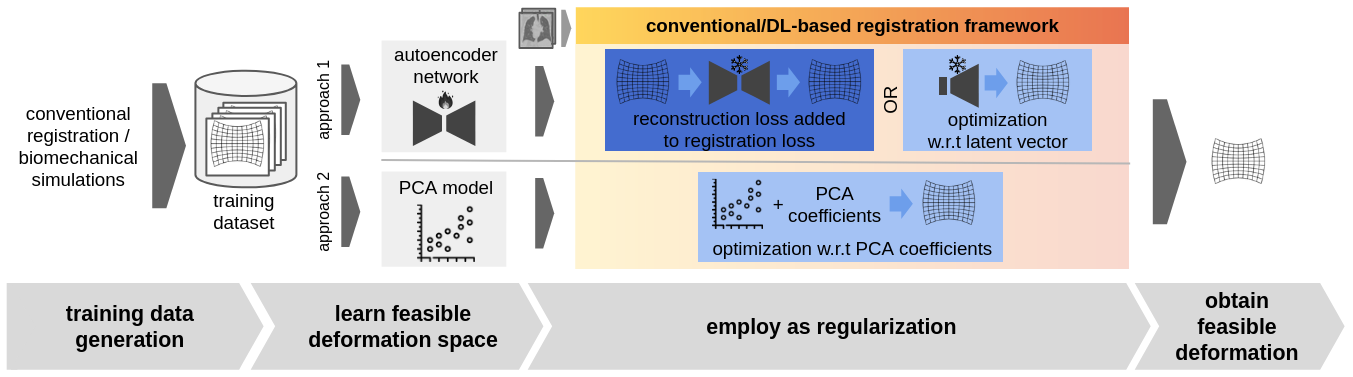}
    \caption{Learned Regularization -- Learned deformation spaces: Low-dimensional deformation spaces representing the set of feasible deformations can be learned from a training dataset. Two approaches of learned deformation spaces are found in the literature: (i) PCA models and (ii) autoencoder networks. Once trained, the regularization model is embedded within the registration framework, either as explicit (\explbox) or implicit (\implbox) regularization.}
    \label{fig:learned}
\end{figure}

\subsubsection{Learned Deformation Spaces} 
Due to the shared characteristics of human anatomy, anatomical structures exhibit similar deformation patterns across different subjects.
Thus, deformations can be assumed to reside within a lower-dimensional subspace. This subspace of feasible deformations can be learned from training data with machine or deep learning models. Once trained, the models offer a compact and efficient deformation space representation with generative capabilities, i.e., they can generate novel deformation instances of the learned space.

The overall process of learning feasible deformation spaces is shown in Fig. \ref{fig:learned}.
Typically, first, a set of ground truth deformations is obtained that adequately spans the feasible deformation space. This training dataset is then used to train the deformation space model. Finally, the trained model is embedded within a registration framework where it parameterizes the transformation. Thus, the registration is implicitly regularized to produce deformations that remain within the learned subspace. This method inherently offers spatially adaptive regularization.

We identify two model types that dominate the literature: 
Learning (i) linear subspaces with principal component analysis (PCA) and (ii) nonlinear manifolds with deep neural networks.

\label{sec:learned-spaces}
\textbf{Principal Component Analysis (PCA)}:
PCA is a linear dimensionality reduction technique that identifies dominant modes of variation in a dataset.
By projecting data onto the learned subspace, PCA enables reconstruction and data generation through linear combinations of the principal eigenvectors.
When applied to deformation datasets, PCA offers statistical deformation models that capture local deformation variations within the training data and (ideally) represent the space of feasible deformations. 

Embedding a trained PCA model within registration greatly reduces the dimensionality of the registration since optimization can be performed with respect to the principal component coefficients instead of the deformation parameters. 
This approach has been employed across various registration techniques, including viscous fluid \citep{wouters2006} and standard gradient-based \citep{hu2008} registration. In the context of diffeomorphic registration, PCA models have been proposed to learn the space of feasible initial momenta in the LDDMM framework \citep{qiu2012} and local velocity fields in the log-Euclidean framework \citep{zeng2018}. 
Additionally, PCA has been applied to control points of FFD B-spline deformations in \cite{loeckx2003}. This further reduces registration complexity.

A challenge with global PCA models is their limited flexibility. To address this, extensions have been developed that enhance representation capabilities. Dual PCA models, for instance, capture complementary deformation properties based on the Jacobian determinants and displacement vectors. Within registration, they ensure that deformations lie in the intersection of both subspaces \citep{xue2006}. A similar approach has been proposed to model tumor-induced deformation and intra-population variations in the context of tumor-to-healthy inter-patient 
registration \citep{mohamed2006}.
Further developments include localized PCA models that represent local patches across the image domain \citep{tang2018}, finding plausible initial solutions with PCA models before registration \citep{kim2008, tanner2009}, and learning a PCA model from a combination of images and segmentation maps \citep{onofrey2015}.

Moreover, learned PCA models can be leveraged within the regularization loss term of the registration. They can avoid infeasible deformations through soft constraints, for example, in a probabilistic formulation \citep{albrecht2008} or by penalizing deviations from the mean of the distribution \citep{berendsen2013}. 
In \cite{cui2017, gu2021}, deformation eigenmodes are constrained to lie within a certain value range for the generation of novel deformations from the PCA model, improving flexibility without sacrificing representational fidelity
Moreover, a deep learning approach using an autoencoder network has been proposed to learn optimal PCA coefficients for a trained PCA model during registration \citep{gao2021}.

Most of the methods above focus on capturing intra-population shape variations for application in inter-patient registration and atlas building.
In contrast, intra-patient registration is often used as a tool \textit{for} the generation of 4D motion models of the lung \citep{he2010, king2012, han2017} or liver \citep{presiwerk2014}. Less work is found that, in turn, applies PCA for regularization purposes \textit{within} registration.
Exceptions include PCA models trained on biomechanical FEM simulations \citep{hu2015} and
FFD-based PCA models trained on pre-to-intra-operative images \citep{onofrey2015}.  
The subject-specific statistical motion model in \citep{jud2017}
is learned from labeled inhale-exhale lung images and successfully transfers sliding motion properties from the training data to unseen, unlabeled images.

\textbf{Deep Neural Networks:}
In contrast to linear PCA models, deep neural networks can learn nonlinear manifolds from training data.
Different model architectures and training strategies have been explored, and we identify three approaches: Employing (i) generative adversarial networks (GANs) that distinguish between ground truth and estimated deformations, (ii) autoencoder networks trained to reconstruct ground truth deformations, and (iii) decoder-only architectures that generate deformations from low-dimensional latent representations.

\textit{GANs} employ generator-discriminator frameworks where the generators act as registration networks. The discriminators are trained to distinguish between estimated and ground truth deformations, which drives the generators to produce results that closely resemble the ground truth data. GANs can be employed to produce biomechanically plausible deformations that align closely with, for instance, biomechanical FEM simulations \cite{hu2018}. 

Another approach involves \textit{autoencoder networks} trained to reconstruct input deformations and, thereby, learning a low-dimensional deformation manifold encoded in the latent space. Trained autoencoders can serve as regularization during registration by penalizing deformations that deviate from the learned manifold. Typically, this is achieved by autoencoding the estimated deformation and adding its reconstruction loss to the registration optimization objective function.
Regularizing autoencoders can be jointly trained with a registration network \cite{bhalodia2019}
or
their training can be decoupled from registration network training, as in \cite{sang2020isbi, sang2021, sang2022}. 
In this case, the autoencoders are trained on separate training datasets and embedded with frozen trained weights in the training of registration networks. Variations of this approach include learning to reconstruct deformation gradients instead of deformations themselves \citep{qin2020} and denoise noisy segmentations \citep{mansilla2020}.
Typically, training data for the autoencoder is generated a priori with conventional registration algorithms \citep{sang2020isbi, sang2021, sang2022}, or biomechanical FEM simulations \citep{qin2020}. 

A third approach uses only the \textit{decoder component} of trained encoder-decoder architectures.
When embedded into a registration framework, optimization is performed directly in the low-dimensional latent space rather than the high-dimensional deformation space. Thus, the registration aims at finding the optimal latent vector that encodes the deformation between $\moving$ and $\fixed$.
\cite{sang2020} have proposed to use alternating backpropagation to jointly optimize the decoder parameters and latent vector.
Similarly, \cite{qin2023} have employed a temporal variational autoencoder that was trained on 4D FEM simulated deformation sequences.
In both methods, the trained network's decoder is integrated into a conventional registration framework, enabling efficient and accurate regularization of deformation properties.

\subsubsection{Learned Test-Time Regularization}
\label{sec:learned-weights}
A core regularization component in the registration objective function (Fig. \ref{eq:opt}) is the regularization weight $\alpha$. It balances the regularization constraints and the similarity loss, and its value determines the strength of the applied regularization.
Typically, $\alpha$ is 
manually tuned as a hyperparameter. However, this is a time-intensive and cumbersome process, particularly for learning-based registration models where separate training is required for each candidate value.
Moreover, a single value applied uniformly across all image pairs is unlikely to yield optimal results due to the variability in deformation properties across different subjects.
To overcome these challenges, advanced deep learning techniques are used within registration networks to learn regularization that can be dynamically adapted at test time.

\begin{figure}[t]
    \centering
    \includegraphics[width=.33\textwidth]{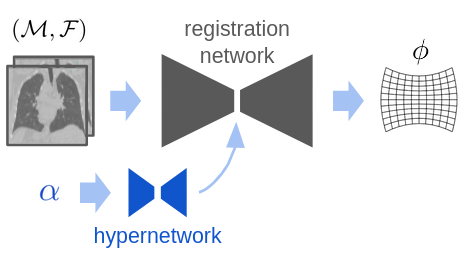}
    \caption{Learned regularization -- Test time regularization: A hypernetwork can allow the user to adapt the regularization weight at test time. During training, the weights are randomly sampled. }
    \label{fig:hypernet}
\end{figure}

A popular approach is to learn the effect of the regularization weight on deformation fields and, thus, on registration network parameters.
During training, $\alpha$ values are randomly sampled to learn a wide range of regularization effects. At test time, the user specifies a desired value, and the registration network estimates a regularized deformation accordingly.
To achieve this, conditional neural networks can be used that employ conditional instance normalization (CIN) layers that normalize and shift feature representations based on the specified $\alpha$ \citep{mok2021conditional}.
Similarly, the HyperMorph framework \citep{hypermorph} leverages a hypernetwork architecture where a secondary network predicts the parameters of the primary registration network, conditioned on $\alpha$ (see Fig. \ref{fig:hypernet}).
Both methods are trained end-to-end and use a global $\alpha$ for balancing a diffusion-based smoothing term. 

Recent advances have extended these methods for spatially adaptive regularization.
\cite{wang2023} enhance the CIN-based approach by introducing a regularization weight matrix for tissue-specific smoothing. 
\cite{chen2023spatially} 
employ a dual decoder registration network that learns the weight matrix, which forms the hypernetwork input alongside the deformation. 
Traditional single-level CIN layers can be extended to multiple levels with a conditional multilevel architecture combining CIN layers, dynamic convolutional layers, and attention layers \cite{zhu2024hierarchical}.
To extend test-time regularization to physics-inspired methods, \cite{reithmeir2024spie} have proposed a hypernetwork that learns the influence of the physical parameters $\lambda, \mu$ on the deformation for linear elastic regularization. This approach has been extended to spatially-adaptive, tissue-specific parameters in \cite{reithmeir2024}.

Apart from offering the user to adapt the regularization at test time, the above methods offer a second advantage: The optimal regularization weight value can be identified at inference time, not only for a test dataset but also for each registration individually. This can be achieved, for instance, with a grid search over the parameter space \citep{hypermorph,reithmeir2024}. A learned approach was instead proposed in \cite{shuaibu2024}. Here, a multilayer perception is trained to predict the Dice score and amount of folding from input images and a specified $\alpha$, enabling a learning-based identification of optimal $\alpha$ values at test time.

Beyond this, automatic adaptation of $\alpha$ during training has been explored to optimize the trade-off between smoothness and registration accuracy during the training process. \cite{xu2022miccai} have introduced a teacher-student architecture where the teacher model updates its weights based on the performance of the student model across consecutive iterations. 
With Monte Carlo dropout, the teacher model estimates deformation and intensity uncertainty, which is then used to dynamically adjust $\alpha$ to each individual image pair and training step. Stronger regularization is applied to more challenging or uncertain image pairs.


\subsubsection{Discussion}
Learned regularization has emerged as a promising approach in medical image registration. Leveraging data-driven techniques to derive local deformation properties from training data offers distinctive advantages over model based and problem specific regularization.
Compared to problem specific methods, fewer explicit modeling and manual tuning towards individual registration instances is required.
Also, a major strength of learned regularization lies in its flexibility. For instance, learned smoothness can automatically find the optimal level of local regularization based on the image information, and over-smoothing can be effectively mitigated \citep{jia2022}.
This automatic adaptability to data is particularly beneficial in scenarios where anatomical structures exhibit highly variable or complex deformation characteristics. 

However, learned regularization also faces some challenges.
In the context of learned deformation spaces, methods require large, high-quality training datasets with accurate ground truth deformations. If the training dataset is smaller than the degrees of freedom of the deformation, PCA models can be overly restrictive and cannot represent the deformation space well \citep{cui2017}. Moreover, since ground truth deformations are inexistent for medical image registration, auxiliary ground truth deformations are commonly generated, for instance, with elastic FEM simulations \citep{hu2008, hu2018, qin2020, qin2023} or conventional registration methods \citep{tang2018, sang2021}. 
Both approaches have inherent limitations. 
Conventional registration-generated training data depend on the size and quality of available training images. 
Learned deformation spaces can only capture deformation properties modeled in the training data, thus requiring adequate design choices during the registration that generates ground truth deformations. For instance, if a global diffusion regularization is used here, the learned deformation spaces can model global smoothness but fail to capture, for instance, local discontinuities. 
In contrast, FEM simulations can generate quasi-unlimited training data but require careful simulation design and predefined physical parameters, which may be ambiguous.

Another challenge for learned deformation spaces is their limited generalizability, particularly when encountering previously unseen anatomical variations or pathological changes.
Recent advancements address this limitation by learning modality-agnostic deformation spaces from segmentation maps rather than images \citep{hu2018,mansilla2020, qin2023}. Additionally, cross-modality transfer has shown promise, where deformation spaces are learned from "easier" modalities, e.g., FBCT or CTA, and successfully transferred to more "complex" modalities, e.g., CBCT or MRI \citep{sang2021, sang2022}.
Interestingly, regularizing PCA models have been widely adopted for representing population-level shape variations in inter-patient registration and atlas building, but their application in intra-patient registration, where they could represent 4D time-dependent motion patterns, remains underexplored.

The innovative advancements of test-time regularization, as introduced in \cite{hypermorph, mok2021conditional}, not only streamline the training process for learning-based registration but also enable scenario- and instance-specific regularization. This aligns well with the inherent variability and subject-specific nature of human body deformations. The tailored regularization during inference bridges the gap between the versatility of instance-specific regularization found in conventional registration and the rapid inference capabilities of learning-based registration. However, training time and model complexity are increased over standard frameworks due to the network architectures and random sampling during training. 

Despite its advantages, learned regularization faces interpretability challenges.
Neural networks inherently function as black-box models, which complicates the task of understanding and verifying the regularization applied during registration.
To better explore learned regularization effects, neural network-based regularization methods can be embedded within conventional registration frameworks where no additional interpretability difficulties are introduced and where it is easier to isolate the regularization effects. This is already found for some of the presented methods, for instance, in \cite{niethammer2019, qin2023}. 
Also, learned regularization still relies on manually designed components. For example, many frameworks predict weight maps for predefined model based regularization terms \citep{mok2021conditional, hypermorph, chen2023spatially, reithmeir2024} or local smoothing filters with fixed isotropic shapes \citep{niethammer2019}.  
Overall, learned regularization often requires significant computational resources and large training data (possibly in addition to registration training data). Careful model design and hyperparameter selection remain essential to prevent overfitting and ensure stable training.

In conclusion, learned regularization is an emerging and highly promising category of regularization methods in medical image registration. While some methods, to some extent, use model based or problem specific regularization, these techniques offer a powerful data-driven approach for capturing complex deformation properties and providing high adaptability of the regularization at inference time.
As the generalizability, robustness, and interpretability of such methods are addressed, learned regularization has the potential to become fundamental to modern medical image registration frameworks.

\section{Open Challenges and Future Perspectives}
\label{sec:discussion}
Regularization is a key component of conventional and learning-based image registration. 
Despite significant advancements in the field, open challenges persist. This section discusses the key limitations of the presented regularization methods and outlines promising directions for future research.

\textbf{Gaps in method transfer from conventional to learning-based registration:}
While many regularization methods have been successfully transferred from conventional to learning-based registration, noticeable gaps remain. 
Most model based regularization methods are widely adopted in learning-based registration. However, physics-inspired regularization has seen limited exploration in deep learning contexts.
Similarly, problem specific regularization methods that deal with locally rigid, cyclic, and sliding motion — once a major focus in conventional registration — are underrepresented in learning-based approaches.

Addressing these gaps could substantially enhance the physical plausibility of learning-based registration models.
The growing research area of physics-informed deep learning presents a valuable opportunity to revisit and expand upon physics-inspired regularization, as demonstrated, for instance, by \cite{ARRATIALOPEZ20235, min2023, min2024}.  Moreover, pre-trained segmentation networks, such as Totalsegmentator \citep{totalsegmentator}, could facilitate the transfer of tissue-specific regularization techniques into learning-based frameworks, potentially leading to more anatomically meaningful results.

\textbf{Overreliance on global smoothness regularization:}
Global $L_2$-norm smoothing is featured in \textit{14 out of 21} methods from the Learn2Reg challenge 2022 \citep{learn2reg2022}, as well as in most state-of-the-art learning-based registration frameworks such as VoxelMorph \citep{balakrishnan2019voxelmorph} or LapIRN \citep{mok2020lapirn}. 
Moreover, the primary focus of many regularization strategies lies in achieving diffeomorphic registrations and minimizing folding. This widespread reliance on global smoothness and diffeomorphic regularization poses a significant limitation in modern registration methods.
It is not only overly restrictive and may fail to capture anatomically plausible deformations, but also might not be suitable in many real-world clinical applications where local discontinuities and pathology-induced topology changes occur. 
Consequently, modern registration frameworks should shift their focus toward flexible, problem specific regularization strategies that can accommodate real-world scenarios and that can handle missing correspondences. 

\textbf{Limited anatomical diversity in the evaluation of problem specific regularization:}
To effectively assess problem specific regularization in medical image registration, the evaluation must consider anatomically appropriate anatomies and scenarios. 
The literature reveals a limited range of anatomical structures for evaluation. It is dominated by the \textit{brain}, which is the most commonly studied anatomy for regularization strategies involving structure specific smoothness, physical properties, multistructure topological consistency, and topological changes.
The \textit{lungs} are the second most studied anatomy, particularly for addressing sliding motion and local rigidity. 
Less explored are cardiac data to evaluate cyclic motion and liver data in the context of sliding motion regularization.
Moreover, brain and cardiac data dominate learned regularization. 

Exceptions include single studies that use private data for evaluation. These special cases involve prostate images in the context of learned deformation spaces \citep{hu2015, hu2018}, images of the colon and cervix in the context of topology changes \citep{suh2011,berendsen2014}, and knee images for various scenarios \citep{shen2019, shen2019region, xu2019, greer2021}. In the context of multistructure registration, exceptions include abdominal, head and neck, and prostate images, e.g., in \cite{freiman2011,greene2009, hu2018weakly, grossbroehmer2024}. However, such data are not used for benchmarking of different methods.

This observed narrow anatomical diversity may result from the \textbf{limited availability of diverse, open-access data}.  
More diverse data that include pathological, longitudinal, and pre-to-post resection images are needed to increase the meaningfulness of method evaluation. Of particular value would be large multiorgan abdominal data, as they align well with many problem specific regularization categories. These include varying organ specific properties, multiorgan topology preservation, local rigidity of bones, sliding motion, and changing topologies within the gastrointestinal tract. The open availability of such data would not only improve the transferability of regularization methods to clinical practice but also facilitate more robust and comparative evaluation across different approaches. 
Overall, more diverse evaluation scenarios will be critical in driving the field forward.

\textbf{Simplified evaluation of regularization properties:}
Evaluating the effectiveness of regularization in image registration remains challenging due to the lack of ground truth deformations. 
Commonly used measures, such as the fraction of negative Jacobian determinants, assess smoothness and related properties such as inverse-consistency, diffeomorphism, and invertibility.
However, these measures alone fail to capture the physical realism of deformations. For instance, sliding motion can increase folding, which leads to higher fractions of negative Jacobian determinants despite being anatomically realistic \citep{heinrich2010}.
In the case that registration involves multistructure segmentations, overlap measures such as the Dice score or Hausdorff distance are often employed -- even though it is well-known that they do not necessarily reflect the registration accuracy \citep{rohlfing2012}. 

More targeted measures have been proposed for specific scenarios. For instance, sliding motion has been evaluated with the maximum shear, calculated via eigenvalue decomposition of the shear tensor \citep{Papiez2014, goksel2016} or the angle between the normal direction and variation in normal motion \citep{schmidrichberg2012}.
In the context of learned deformation spaces, the reconstruction error provides a measure of deformation plausibility.
Yet, such targeted metrics are rarely employed in evaluating registration frameworks that incorporate problem specific or learned regularization.
Overall, there is a strong need to develop more sophisticated measures to assess complex and local deformation properties, for instance, for evaluating local rigidity and topological changes.

\textbf{Translation of problem specific regularization to clinical practice is hindered:}
Despite the variety of regularization methods developed for specific registration tasks, their translation to clinical practice faces several challenges.
In addition to the frequent reliance on global smoothness regularization, as discussed above, simplified conditions in problem specific regularization cause difficulties and require careful consideration.
Examples include the assumption of linear sliding boundaries, as in \cite{schmidrichberg2012, delmon2013}, and single rigid regions, as in \cite{Lester1999}. Such assumptions limit their applicability to more complex real-world scenarios.

Additionally, the inherently \textbf{subject- and instance-specific} nature of registration problems poses a significant hurdle for generalizing problem specific regularization techniques across different patients. 
For instance, cyclic constraints are not suitable for patients with arrhythmia \citep{wiputra2020}, and physical properties such as tissue elasticity can vary due to individual factors such as age \citep{cocciolone2018elastin}. 
These challenges highlight the importance of robust and generalizable regularization methods that can adapt to individual registration problems and they show that the successful application of problem specific regularization remains in the hands of the user.\\

Given the rapid advancements of deep learning research, we anticipate further developments in methods that can precisely and automatically tailor regularization to individual image pairs and leverage advanced deep learning architectures to capture feasible and local deformation properties more effectively. 
Recent innovations, such as the conditional regularization approaches proposed by \cite{hypermorph, mok2021conditional}, have introduced a paradigm shift in learning-based registration towards adapting the regularization to user requirements at test time while offering high registration speed. 
Overall, user interaction for adaptive regularization at test time could gain importance in the future.

Another promising direction involves exploiting advanced deep learning architectures for regularization purposes, as done with physics-informed networks (PINNs) and Lipschitz-continuous ResNet blocks to solve biomechanics-inspired and diffeomorphic PDEs \citep{joshi2022, ARRATIALOPEZ20235, min2023}, or with attention mechanisms to provide targeted control over local regions during registration \citep{feng2024}. 
Beyond this, fully automated adaptation of regularization to given registration instances represents a compelling direction for future research.
An already versatile smoothness regularization is offered by the GradICON regularization \citep{tian2022} which has demonstrated robust smoothness across diverse datasets without requiring customized regularization strategies. It is successfully leveraged in the recently introduced UniGradICON framework \citep{tian2024unigradicon}, presented as the first foundation model for image registration.
Further advanced approaches, such as Auto-ML frameworks \citep{automl} for automatic regularization learning or data-driven general purpose regularization, remain to be explored.

\section{Conclusion}
Regularization is a fundamental building block of image registration. 
It ensures that derived deformations align with physical and anatomical plausibility.
This review systematically classified the wide range of proposed regularization methods in the literature. Three main categories were identified: (i) \textit{model based regularization} that uses prior assumptions, (ii) \textit{problem specific regularization} that incorporates prior data knowledge, and (iii) \textit{learned regularization} that derives deformation properties from training data. 
Each category addresses distinct challenges in registration while contributing to the ongoing advancements of medical image registration.

Model based regularization remains the foundation of both conventional and learning-based registration algorithms.
Techniques such as $L_2$-norm smoothness and diffeomorphic registration offer robust and interpretable frameworks, which have been effectively integrated into registration networks, for instance, through differentiable regularization layers and explicit training loss terms. 
However, while their handcrafted nature facilitates interpretability, the global assumptions often limit their ability to model complex and heterogeneous deformations seen in clinical data.

Problem specific regularization addresses these limitations by leveraging spatial and contextual knowledge to enable more realistic deformation modeling. By locally adapting the deformation properties, these methods are suitable for challenging scenarios, including sliding motion, topological changes, and organ specific deformation properties. Despite their potential, their dependency on high quality spatial information and their design toward specific registration problems can restrict their scalability and generalizability, particularly in learning-based registration. 
While many problem specific regularization methods have already been successfully adapted to learning-based frameworks, some remain to be transferred.

Learned regularization leverages data-driven techniques to learn deformation properties from a training dataset, enabling more flexible solutions. The parameterization of the regularization as neural networks opens new opportunities, such as learning low-dimensional feasible deformation spaces and exploiting the inherent properties of advanced architectures for regularization purposes. A key innovation is test time regularization within learning-based registration frameworks, which offers adaptation to subject- and instance-specific deformation properties of human body motion. An open challenge is posed by the limited interpretability and need for additional training datasets compared to model based and problem specific methods. 

Looking ahead, hybrid regularization methods that combine the interpretability and robustness of model based methods with the adaptability of learned techniques are particularly promising. Additionally, fully automated regularization adaptation and general purpose methods may gain importance in the future.
We want to emphasize the strong need for (i) more diverse open-access registration datasets that represent a broader spectrum of anatomical and pathological conditions, and (ii) improved evaluation measures that assess local deformation properties, which are equally critical for driving the field forward.
By addressing the presented research gaps, regularization can play an even greater role in improving the accuracy, reliability, and applicability of image registration in clinical practice.

This review has highlighted the importance of regularization, which persists today in the era of learning-based image registration, and demonstrated that it is a rapidly evolving research field -- alongside the development of novel modern registration algorithms. Advancements in regularization methods for image registration could also impact connected research fields, including motion correction, segmentation, and medical image reconstruction. 
We hope that this review inspires the research community to reconsider regularization strategies in current state-of-the-art registration methods and to explore open challenges and novel developments to further advance the field of image registration methods.

\bibliographystyle{cas-model2-names}

\bibliography{bib, data_driven, data_specific, sliding_motion}

\end{document}